\documentclass[longauth]{aa} % for the long lists of affiliations 

\usepackage{graphicx}
\usepackage{txfonts}
\usepackage{fixltx2e}
\usepackage{xcolor}
\usepackage{lineno}
\newcommand*\patchAmsMathEnvironmentForLineno[1]{%
  \expandafter\let\csname old#1\expandafter\endcsname\csname #1\endcsname
  \expandafter\let\csname oldend#1\expandafter\endcsname\csname end#1\endcsname
  \renewenvironment{#1}%
     {\linenomath\csname old#1\endcsname}%
     {\csname oldend#1\endcsname\endlinenomath}}%
\newcommand*\patchBothAmsMathEnvironmentsForLineno[1]{%
  \patchAmsMathEnvironmentForLineno{#1}%
  \patchAmsMathEnvironmentForLineno{#1*}}%
\AtBeginDocument{%
\patchBothAmsMathEnvironmentsForLineno{equation}%
\patchBothAmsMathEnvironmentsForLineno{align}%
\patchBothAmsMathEnvironmentsForLineno{flalign}%
\patchBothAmsMathEnvironmentsForLineno{alignat}%
\patchBothAmsMathEnvironmentsForLineno{gather}%
\patchBothAmsMathEnvironmentsForLineno{multline}%
}

\usepackage{hyperref}
\usepackage{breakurl}

\newcommand{\fermi}{\textit{Fermi}-LAT}
\newcommand{\swift}{\textit{Swift}-XRT}
\newcommand{\uvot}{\textit{Swift}-UVOT}
\newcommand{\src}{QSO B1420+326}
\newcommand{\nodata}{\multicolumn{1}{c|}{--}}
\newcommand{\nodataa}{\multicolumn{1}{c}{--}}

\makeatletter
\renewcommand*\aa@pageof{, page \thepage{} of \pageref*{LastPage}}
\makeatother
\usepackage{graphicx}	% Including figure files
\usepackage{amsmath}	% Advanced maths commands
\usepackage{amssymb}	% Extra maths symbols

\titlerunning{MWL campaign on \src}
\begin{document}
\title{VHE gamma-ray detection of FSRQ \src\ and modeling of its enhanced broadband state in 2020}

\authorrunning{MAGIC Collaboration et al.}

% authors 04.09.2020  Format AA
%commento di prova file normale%
\author{
\normalsize
V.~A.~Acciari\inst{1} \and
S.~Ansoldi\inst{2} \and
L.~A.~Antonelli\inst{3} \and
A.~Arbet Engels\inst{4} \and
M.~Artero\inst{5} \and
K.~Asano\inst{6} \and
D.~Baack\inst{7} \and
A.~Babi\'c\inst{8} \and
A.~Baquero\inst{9} \and
U.~Barres de Almeida\inst{10}
% ...
}
\institute { Inst. de Astrof\'isica de Canarias, E-38200 La Laguna, and Universidad de La Laguna, Dpto. Astrof\'isica, E-38206 La Laguna, Tenerife, Spain
\and Universit\`a di Udine and INFN Trieste, I-33100 Udine, Italy
\and National Institute for Astrophysics (INAF), I-00136 Rome, Italy
\and ETH Z\"urich, CH-8093 Z\"urich, Switzerland
\and Institut de F\'isica d'Altes Energies (IFAE), The Barcelona Institute of Science and Technology (BIST), E-08193 Bellaterra (Barcelona), Spain
\and Japanese MAGIC Group: Institute for Cosmic Ray Research (ICRR), The University of Tokyo, Kashiwa, 277-8582 Chiba, Japan
\and Technische Universit\"at Dortmund, D-44221 Dortmund, Germany
\and Croatian MAGIC Group: University of Zagreb, Faculty of Electrical Engineering and Computing (FER), 10000 Zagreb, Croatia
\and IPARCOS Institute and EMFTEL Department, Universidad Complutense de Madrid, E-28040 Madrid, Spain
\and Centro Brasileiro de Pesquisas F\'isicas (CBPF), 22290-180 URCA, Rio de Janeiro (RJ), Brazil
% ...
}
\vspace{-1.5cm}
\offprints{
contact.magic@mpp.mpg.de, \\
$^*$Corresponding authors: J.~Sitarek, F.~D'Ammando, R.~Angioni, S.~Nozaki\\
$^\dagger$ deceased
}
\date{Received ; accepted }

% Abstract of the paper (<=250 words)
\abstract
% context  (optional)
{\src\ is a blazar classified as a Flat Spectrum Radio Quasar (FSRQ). 
  In the beginning of 2020 it underwent an enhanced flux state. 
An extensive multiwavelength campaign allowed us to trace the evolution of the flare. 
}
% aims
{  We search for VHE gamma-ray emission from \src\ during this flaring state. 
We aim to characterize and model the broadband emission of the source over different phases of the flare.
}
% methods
{The source was observed with a number of instruments in radio, near infrared, optical (including polarimetry and spectroscopy), ultraviolet, X-ray and gamma-ray bands. 
We use dedicated optical spectroscopy results to estimate the accretion disk and the dust torus luminosity. 
We perform spectral energy distribution modeling in the framework of combined  Synchrotron-Self-Compton and External Compton scenario in which the electron energy distribution is partially determined from acceleration and cooling processes. }
% results
{
During the enhanced state the flux of both SED components drastically increased and the peaks were shifted to higher energies.  Follow up observations with the MAGIC telescopes led to the detection of very-high-energy gamma-ray emission from this source, making it one of only a handful of FSRQs known in this energy range. 
Modeling allows us to constrain the evolution of the magnetic field and electron energy distribution in the emission region. 
The gamma-ray flare was accompanied by a rotation of the optical polarization vector during a low polarization state. 
Also, a new, superluminal radio knot contemporaneously appeared in the radio image of the jet.
The optical spectroscopy shows a prominent FeII bump with flux evolving together with the continuum emission and a MgII line with varying equivalent width. \vspace{0.5cm}
}
% conclusions (optional)
{}


%%%%%%%%%%%%%%%%%%%%%%%%%%%%%%%%%%%%%%%%%%%%%%%%%%
%%%%%%%%%%%%%%%%% BODY OF PAPER %%%%%%%%%%%%%%%%%%

\maketitle

\section{Introduction}

\src , also known as OQ 334, is a blazar located at redshift of 0.682 \citep{hw10}. 
Based on its radio spectrum it has been classified as a Flat Spectrum Radio Quasar \citep[FSRQ;][]{he07}.
BL Lac objects are divided between low-, intermediate- and high- synchrotron peaked (LSP, ISP, HSP), while FSRQs are usually only LSP objects. 
At HE (GeV) energies FSRQs populate the majority of the extragalactic gamma-ray sky. 
Among the associated blazars in the \fermi{} Fourth Source Catalog \citep[4FGL;][]{abd20} there are 694 FSRQs compared to 1131 BL Lac objects.  
In the very-high-energy (VHE, $\gtrsim 100$\, GeV) range despite over 60 BL Lac objects have been detected by Imaging Atmospheric Cherenkov Telescopes (IACTs), only about 8 FSRQs are known to emit in this energy range\footnote{See \burl{http://tevcat.uchicago.edu/}, in case of some sources the classification as FSRQ or BL Lac is however uncertain.}. 
There are a few probable reasons that contribute to the difference between the  number of blazars detected at HE and VHE. 
The peak of the gamma-ray emission in the Spectral Energy Distribution (SED) of FSRQs is usually shifted to lower energies compared to BL Lac objects (see, e.g., \citealp{gh16})
%\footnote{
%The synchrotron peak of FSRQ objects is also usually at lower frequency than for BL Lac objects, namely 
Also, some of the source may have enhanced internal absorption in the radiation field of the Broad Line Region (BLR, see, e.g., \citealp{lb06}, but note also \citealp{co18}) via $e^+e^-$ pair production process. 
FSRQs are considered to be more luminous sources than BL Lacs, which permits them to be detected at larger cosmological distances. 
However, for sources located at high redshift ($z\sim 1$), the VHE gamma-ray part of the spectrum is severely absorbed in the pair production process on Extragalactic Background Light (EBL, see, e.g., \citealp{do11}), hampering the discovery potential in this energy range. 
Softer gamma-ray spectra resulting from both effects further make detection of FSRQs with IACTs more difficult. 
Finally, FSRQs are known to be extremely variable (see, e.g., \citealp{me19}), which is another complication in observing those sources with instruments with relatively small fields of view such as IACTs. 
The VHE gamma-ray flux has been seen to vary even by two orders of magnitude (see, e.g., \citealp{dammando19, za19}). 
Emission variability has been observed down to a time scale of $\sim10$ minutes \citep{al11}. 
Due to the strong variability, currently the most successful approach for studying the VHE gamma-ray emission of FSRQs is a follow up of alerts of enhanced activity at lower frequencies. 
To date all the cases of discovery of VHE gamma-ray emission from FSRQs have occurred either during short flaring activities or longer high states. 
A notable counterexample is PKS\,1510--089, from which persistent VHE gamma-ray emission has been observed during a low flux level at HE \citep{acc18}. 
Since the number of known VHE FSRQs is still very small, it is important to observe those objects to look for common patterns and differences in their emission and to investigate if the same processes are responsible. 
Moreover, the observations need to be multiwavelength and thus contemporaneously cover the broad energy range of the spectrum, which is often difficult to achieve due to fast variability. 

\src\ is known to be strongly variable, in particular in  the HE range\footnote{\burl{https://fermi.gsfc.nasa.gov/ssc/data/access/lat/msl_lc/source/OQ_334}}.
A few periods of HE flux enhancement from the source have been observed by  \fermi\ so far, the most recent one starting in 2019 December \citep{cc20}. 
The high state continued, and based on the HE flux enhancement, the MAGIC telescopes have performed follow up observations and discovered VHE gamma-ray emission from \src\ in 2020 January \citep{mi20}. 
The source is the fourth most distant VHE gamma-ray source known. 

Thanks to the duration of the enhanced state we were able to alert other observatories and trace the development of the flare in a broad range of wavelengths, from radio up to VHE gamma rays \citep[e.g.,][]{minev20,dammando20,ramazani20}. 
Since contemporaneously to gamma-ray flaring activity in FSRQs, often ejection of new knots in their jets have been reported (see, e.g., \citealp{al14, li15, ra18}), we also organized VLBI (very-long-baseline radio interferometry) observations of the source following the high state.

In this paper we report on the broadband observations of \src\ triggered by the 2019/2020 high state and other contemporaneous multiwavelength (MWL) data and on their interpretation.
In Section~\ref{sec:obs} we describe the instruments involved in the campaign, the data taken and the analysis methods.
The results of the observations are reported in Section~\ref{sec:res}.
In Section~\ref{sec:model} we model the broadband emission of the source in different phases of the high state.
Results are summarized in Section~\ref{sec:conc}.

%We use cosmological parameters $H_0$=70~km/s/Mpc, $\Omega_\Lambda$ = 0.7, and  $\Omega_M$ = 0.3. 
We use cosmological parameters $H_0$=67.4~km/s/Mpc, $\Omega_\Lambda$ = 0.6847, and  $\Omega_M$ = 0.315 \citep{planck18}. 

\section{Observations and data analysis} \label{sec:obs}
\src\ was observed in 2020 January/February in a broad energy range by a number of instruments that either monitor the source, or had responded to a target of opportunity (ToO) announcement for a high state of the blazar, including also publicly released results. 
We report on the observations performed in 
radio (VLBA, OVRO, Mets\"ahovi, Badary RT32),
NIR (CANICA),
optical polarization and photometry (Perkins, LX-200, AZT-8+ST7), 
optical photometry  (Siena and Rozhen Observatories, ASAS-SN monitoring, MIRO/MFOSC-P, REM), 
optical spectroscopy (LDT), 
optical and UV from space satellites (\uvot\ and \textit{XMM}-OM), 
X-rays (\swift\ and {\it XMM-Newton}),
HE gamma rays (\fermi )
and VHE gamma rays (MAGIC). 
To put the results in the context of earlier measurements we also use archival data retrieved via the Space Science Data Center\footnote{\burl{https://www.ssdc.asi.it/}} from the  catalogs: GB6 \citep{ssdc_gb6}, FIRST \citep{ssdc_first}, NVSS \citep{ssdc_nvss}, CLASS \citep{ssdc_class}, JVASPOL \citep{ssdc_jvaspol}, WISE \citep{ssdc_wise}, GALEX \citep{ssdc_galex}, Planck \citep{ssdc_planck}, 2RXS \citep{ssdc_2rxs}, SDSS \citep{ssdc_sdss}.
%\cite{ssdc_gb6, ssdc_first, ssdc_nvss, ssdc_rass, ssdc_class, ssdc_jvaspol, ssdc_wise, ssdc_galex, ssdc_planck, ssdc_2rxs,ssdc_sdss}.  
% removed ssdc_3fgl
In the archival data sample we also include the lowest and the highest \swift{} states of the source (MJD 57631 and 58831), and the low state observed by \fermi{} (integrated from the mission start, MJD 54683 until MJD 57754).

\subsection{MAGIC}
 
MAGIC is a system of two IACTs with a mirror dish diameter of 17\,m each \citep{al16a}. 
The telescopes are located in the Canary Islands, on La Palma ($28.7^\circ$\,N, $17.9^\circ$\,W), at a height of 2200 m above sea level. 

\src\ has been observed by MAGIC on a few occasions following enhanced states at lower energies. 
We report on observations between 2019 December 31 (MJD=58848) and 2020 February 6 (MJD=58885). 
The observations consisted of several triggers of MAGIC Target of Opportunity program and are therefore, also due to bad weather periods, not continuous. 
The data selection was based on the atmospheric transmission and rates of background events.  
The total amount of good quality data is 14.0\,hr, out of which 2.9\,hr, taken between  2019 December 31 (MJD=58848) and 2020 January 4 (MJD=58852), has been taken with a special low-energy analogue trigger: SUM-Trigger-II \citep{ga14}. % 13.96
The data are analyzed using MARS, the standard analysis package of MAGIC \citep{za13, al16b}.
The data selection is based mainly on the atmospheric transmission measured with LIDAR \citep{fg15} and on hadronic background rates.
The effect of atmospheric absorption is corrected using LIDAR information.  
The Sum-Trigger-II part of the dataset is analyzed with dedicated low-energy analysis procedures including a special image cleaning (the so-called MaTaJu cleaning) with the cleaning thresholds tuned to the extragalactic field of view of \src{} \citep{sh13,ce19}. 
For the part of the dataset during which no signal is detected, upper limits on the flux are computed following \cite{ro05} at 95\% confidence level. 

\subsection{\fermi}

The Large Area Telescope (LAT) is a pair-conversion telescope, launched on 2008 June 11 (MJD=54628) as one of the two scientific instruments on board the \textit{Fermi Gamma-ray Space Telescope} \citep{atw09}. 
%Its energy range extends down to 20 MeV and up to $>1$ TeV \citep{aje17}, with peak sensitivity in the range 0.3--100\,GeV. 
Its energy range extends down to $\sim$ 30 MeV and up to $\sim300$ GeV, with peak sensitivity at $\sim 1$\,GeV. %\citep{atw09}
In the 4FGL~\citep{abd20}, \src\ is associated to the gamma-ray source 4FGL\,J1422.3+3223, which has a $>100$ MeV flux of $(9.1\pm1.3)\times10^{-9}$ \,cm$^{-2}$\,s$^{-1}$ and a power-law spectrum with photon index $2.38\pm0.07$, obtained from data between 2008 June 11 (MJD=54682.65) and 2016 August 2 (MJD=57602.24). \src{} does not appear in any of the \textit{Fermi}-LAT hard-spectrum source catalogs~\citep[see, e.g., the 3FHL catalog,][]{aje17}, which is consistent with its relatively steep spectrum.

We use the Python package \texttt{Fermipy} \citep{woo17} to analyze the \textit{Fermi}-LAT data. We use Pass8 event data~\citep{atw13} and select photons of the \texttt{SOURCE} class, in a square region of interest (ROI) of 10$^\circ\times10^\circ$, centered at the position of the target source. 
We perform a binned analysis with 10 bins per decade in energy and 0.1$^{\circ}$ binning in RA and Dec, in the energy range 0.1--300 GeV, adopting the instrument response functions \texttt{P8R3\_SOURCE\_V2}. A correction for energy dispersion is included for all sources in the model except for the isotropic diffuse components. We apply a cut to include only the gamma-rays with zenith angle $<90^{\circ}$ to  limit contamination from the Earth's limb. We include in the model of the region all sources listed in the 4FGL within 15$^{\circ}$ from the ROI center, along with the Galactic \citep{ace16} and isotropic diffuse emission models (\href{https://fermi.gsfc.nasa.gov/ssc/data/analysis/software/aux/4fgl/gll\_iem\_v07.fits}{\texttt{gll\_iem\_v06.fits}} and \href{https://fermi.gsfc.nasa.gov/ssc/data/analysis/software/aux/4fgl/iso\_P8R3\_SOURCE\_V2\_v1.txt}{\texttt{iso\_P8R3\_SOURCE\_V2\_v1.txt}}, respectively). 

%We first perform a likelihood fit using the full \textit{Fermi}-LAT data set available at the time of the analysis, including events in the time range 2008 August 4, 15:43:36.0 UTC to 2020 February 11, 00:00:00.0 UTC (MJD 54682.66--58890.00). 
We first perform a likelihood fit using the full \textit{Fermi}-LAT data set available at the time of the analysis, including events in the time range 2008 August 4 to 2020 February 11 (MJD 54682.66--58890.00). 
We fit the full spectrum of the target source, the diffuse models, and the normalization of catalog sources within 5$^{\circ}$ as free parameters. We also optimize the target source localization, 
taking advantage of the $\sim3.5$ extra years of data with respect to 4FGL. The detection significance is estimated with the Test Statistic \citep[TS,][]{mat96}.
We search for new sources by performing a TS map of the ROI. No significant (TS$>$25) new gamma-ray source is found in this analysis. Although mild excesses with TS$\sim$10 are seen in the residuals, such fluctuations are to be expected when periods over long time ranges such as this are considered, and therefore we choose not to add these excesses as point sources in the ROI model. The model resulting from this initial fit is used as an input for the computation of the HE gamma-ray light curves.

The light curves are calculated by performing a likelihood fit in each time bin. The fitting strategy is designed to adjust the number of free parameters to the photon statistics available in each bin. In the \textit{Fermi}-LAT source catalogs, a source is considered detected if TS is at least 25. 
In each light curve bin, we fit the full spectrum of the target source and the normalizations of the sources within 3$^{\circ}$ from the ROI center. 
If the target source has TS $<$ 25, we progressively restrict the free parameters in the fit, reloading the average model at each step. 
First, we reduce the sources with free normalization to a radius of 1$^{\circ}$, then we freeze all sources except the target, and finally, if the target is not significantly detected, we only fit its normalization, leaving the spectral parameters fixed to the average value from the initial model. We consider the target source to be detected in a given time bin if TS $>9$, and the signal-to-noise ratio (that is, flux divided by its error, or $F/\Delta F$) in that bin is larger than two. If this is not the case, we reported a 95\% confidence upper limit on the flux.

We report light curves with fixed binning of one day and 30 days, and one with adaptive binning \citep{lot12}, setting a constant relative flux uncertainty of 15\%. % and using an optimal energy of $\sim230$\,MeV. 
The latter method provides an estimate of the shortest time scale over which it is possible to obtain a statistically significant detection and a robust determination of the target's spectral parameters.

We calculate 0.1--300 GeV spectral energy distributions (SEDs) in the time intervals listed in Section~\ref{sec:res}, by performing a likelihood fit in several energy bins. %Table~\ref{tab:periods}
The number of bins is optimized as a trade-off between energy resolution and photon statistics. We also perform an analysis including data from 2008 August 4 (MJD=54682) to 2017 January 1 (MJD=57754), to compute a quiescent-state \textit{Fermi}-LAT SED to which the high-state ones could be compared. This time range was chosen based on the monthly light curve, which shows some early signs of flaring activity between the second half of 2017 and the first half of 2018 (see Appendix~\ref{app:longterm}). %Fig.~\ref{fig:mwl_lc_longterm}
This time range is quite similar to the one corresponding to the 4FGL catalog, but provides a small increase in photon statistics due to the later end time.

Finally, we calculate the probability for each single gamma ray recorded by the \textit{Fermi}-LAT of being associated with \src{}, using the \texttt{gtsrcprob} tool, in order to investigate the highest energy photons in the 0.1-300\,GeV energy range.

\subsection{X-ray}

The {\em Neil Gehrels Swift Observatory} \citep{gehrels04} carried out 26 observations of the source between 2018 January 20 (MJD=58138) and 2020 February 10 (MJD=58889), in particular on 14 individual days between 2020 January 2 and February 10 (MJD=58850 -- 58889). 
The pointed observations were performed with both the X-ray Telescope \citep[XRT;][0.2--10.0 keV]{burrows05} and the Ultraviolet/Optical Telescope \citep[UVOT;][170--600nm]{roming05}. 
The hard X-ray flux of this source is below the sensitivity of the BAT instrument for the short exposures of these observations; therefore, the data from this instrument are not used. 
Moreover, the source is not present in the {\em Swift} BAT 105-month catalog \citep{oh18}.

All XRT observations were performed in photon counting (PC) mode. The XRT spectra are generated with the {\em Swift} XRT data products generator tool at the UK Swift Science Data Centre\footnote{\burl{http://www.swift.ac.uk/user_objects}} \citep[for details see][]{evans09}. Ancillary response files are generated with \texttt{xrtmkarf}, and account for different extraction regions, vignetting and point-spread function corrections. We use the spectral redistribution matrices v014 in the Calibration data base maintained by \texttt{HEASARC}. Some of the spectra have very few photons, and so we are not able to use $\chi^{2}$ statistics. To maintain the homogeneity in our analysis, we grouped the obtained spectra using the task \texttt{grppha} to have at least one count per bin and we perform the fit with the Cash statistics \citep{cash79}. The data collected during 2019 June 27 and 29 (MJD=58661 and 58663) are summed in order to have enough statistics to obtain a good spectral fit.
We fit the spectra with an absorbed power-law using the photoelectric absorption model \texttt{tbabs} \citep{wilms00}, with a neutral hydrogen column density fixed to its Galactic value \citep[$1.08 \times 10^{20}$ cm$^{-2}$;][]{bekhti16}. 
We apply also a log-parabola model to the XRT data, testing if this model is preferred over a single power-law model on a statistical basis by applying an F-test. The log-parabola model is preferred over a simple power-law model only on 2020 January 28 and at 95\% confidence level. However, this can be due to the low statistics of the single XRT observations.

{\em XMM-Newton} \citep{jansen01} observed the source on 2020 January  24 between 04:44:07 and 11:49:07 (MJD 58872.3 -- 58872.5) for a total duration of 25 ks (observation ID 0850180201). 
The 3 EPIC cameras were operated in the large-window mode with medium filter. The data are reduced using the {\em XMM-Newton} Science Analysis System ({\small SAS v16.0.0}), applying standard event selection and filtering. Inspection of the background light curves show that no strong flares were present during the observation, with good exposure times of 20, 24 and 24 ks for the pn, MOS1 and MOS2, respectively. For each of the detectors the source spectrum is extracted from a circular region of radius 30 arcsec centered on the source, and the background spectrum from a nearby region of radius 30 arcsec on the same chip. All the spectra are binned to contain at least 20 counts per bin to allow for $\chi^2$-based spectral fitting. All spectral fits are performed over the 0.3--10~keV energy range using {\small XSPEC v.12.10.1}. The energies of spectral features are quoted in the source rest frame. 
All errors are given at the 90\% confidence level.
The data from the three EPIC cameras are initially fitted separately, but since good agreement is found  ($<5\%$) we
proceed to fit them together. Galactic absorption is included in all fits using the \texttt{tbabs} model.

Three different models are applied: a simple power-law, a broken power-law, and a log-parabola model. 
The results of the fits are presented in Table~\ref{xmmfits}. The F-test shows an improvement of the fit using both a broken power-law and a log-parabola model with respect to the simple power-law, with the log-parabola model providing the best fit. 

In order to check for the presence of intrinsic absorption, a neutral absorber at the redshift of the source is added to this model, but it does not improve the fit quality and thus is not required. Moreover, no Fe line was detected in the spectrum. The 90\% upper limit on the equivalent width (EW) of a narrow emission line at 6.4~keV is EW$<10$~eV.

\begin{table}
\begin{center}
\begin{tabular}{lll}
\hline \\ [-4pt]
Model & Parameter & Value \\[4pt]  
\hline\\[-6pt]
Power-law & $\Gamma$ & $1.95\pm 0.01$  \\
      & Flux (0.3--10 keV)  &  $(5.2\pm 0.1) \times 10^{-12}$     \\
      &  $\chi^2/\rm{d.o.f.}$          &   1627.79/379  \\
\hline 
Broken Power-law & $\Gamma_1$ & $2.38 \pm 0.05$  \\
      & $E_{\rm\,break}$ & $1.22^{+0.10}_{-0.08}$  \\
      & $\Gamma_2$ & $1.58 \pm 0.04$  \\
      & Flux (0.3--10 keV) &  $(6.0\pm 0.1) \times 10^{-12}$     \\
      &  $\chi^2/\rm{d.o.f.}$          &   472.21/377  \\
\hline
Log Parabola & $\alpha$ & $2.12 \pm 0.01$  \\
             & $\beta$  & -0.60 $\pm$ 0.03 \\
             & Flux (0.3--10 keV) &  $(6.2\pm 0.1) \times 10^{-12}$     \\
      &  $\chi^2/\rm{d.o.f.}$          &   432.42/378  \\
\hline
\end{tabular}
\end{center}
\caption{Summary of fits to the 0.3--10~keV {\em XMM-Newton} spectrum of the source. Fits also include absorption fixed at the Galactic value. Flux and E$_{\rm\,break}$ (in the source rest frame) are given in units of erg cm$^{-2}$ s$^{-1}$ and keV, respectively.}\label{xmmfits}
\end{table}
 
\subsection{Optical and UV from space-based telescopes}

During the {\em Swift} pointings, the UVOT instrument observed \src\ in all its optical ($v$, $b$ and $u$) and UV ($w1$, $m2$ and $w2$) photometric bands \citep{poole08,breeveld10}. For each epoch, possible multiple exposures in the same filter are first summed with the task \texttt{uvotimsum} and then analyzed using the \texttt{uvotsource} task included in the \texttt{HEAsoft} package (v6.28) with the 20201026 release of the Swift/UVOTA CALDB. %including the new UVOT sensitivity calibration file to account for the loss of sensitivity with time in the UV and white filters after 2017.
We check if the observations are affected by small-scale sensitivity problems\footnote{https://swift.gsfc.nasa.gov/analysis/uvot\_digest/sss\_check.html}.
Source counts are extracted from a circular region of 5 arcsec radius centered on the source, while background counts are derived from a circular region of 20 arcsec radius in a nearby source-free region. 
%The calibration is done with \texttt{swusenscorr20041120v006.fits} file released in 2020 September. 

The Optical Monitor (OM) on board of the {\em XMM-Newton} satellite observed the source in the $u$, $w1$, $m2$, and $w2$ filters in imaging mode together with a fast readout window. The total exposure times of the imaging observations are: 3500\,s ($u$), 3500\,s ($w1$), 4400\,s ($m2$) and 4400\,s ($w2$). The data are processed using the SAS tasks \texttt{omichain} and \texttt{omfchain}.

The UVOT and OM flux densities are corrected for extinction using the E(B--V) value of 0.010 from \citet{sf11} and the extinction laws from \citet{cardelli89}.

\subsection{Optical from ground-based telescopes}
% 

%REM
The REM telescope \citep{zerbi01, covino04}, a robotic telescope located at the ESO Cerro La Silla observatory (Chile), performed optical photometric observations of \src\ in the period 2020 January 24--February 6 (MJD=58872 -- 58885). Observations were carried out with the Optical Slitless Spectrograph (ROSS2) obtaining three 240 s integration images in the optical g$^{\prime}$, r$^{\prime}$, i$^{\prime}$ bands. The REM data presented here were obtained as ToO observations triggered by the high gamma-ray flux observed by {\em Fermi}-LAT after the MAGIC detection. 
Instrumental magnitudes are obtained via aperture photometry and absolute calibration is performed by means of secondary standard stars in the field reported by the AAVSO Photometric All-Sky Survey (APASS) catalog\footnote{\burl{https://www.aavso.org/apass}}. Transformation between the u$^{\prime}$\,g$^{\prime}$\,r$^{\prime}$\,i$^{\prime}$\,z$^{\prime}$ and UBVRI photometric systems are performed using the equations reported in \citet{jester05}\footnote{\burl{https://www.sdss.org/dr16/algorithms/sdssUBVRITransform/}}. 

Optical photometric (BVRI) and polarimetric (R band) observations were carried out at the 1.83~m Perkins telescope (Flagstaff, AZ, USA), 40~cm LX-200 telescope (St.Petersburg, Russia) and 70~cm AZT-8 telescope (Crimea) from 2020 January 23 to April 21 (MJD 58871 -- 58960).
The photometric data are reduced using differential aperture photometry with respect to comparison stars in the quasar field\footnote{see \burl{https://vo.astro.spbu.ru/vlar/opt_thumbs/b21420_1.png}}. 
The polarimetric observations obtained at the Perkins telescope were performed and reduced in the same manner as described in \cite{Jorstad2010}. The details of polarimetric observations carried out at the LX-200 and AZT-8 telescopes can be found in \cite{Larionov2008}. 

% MIRO
Photometric optical observations of \src\ were carried out with the MFOSC-P instrument \citep{sr18}, used in imaging mode, mounted on the 1.2 m telescope of Mount Abu IR Observatory (MIRO\footnote{\burl{https://www.prl.res.in/~miro/}}). 
The source was observed in B, V, R, and I bands (Johnson-Cousins filters) on 2020 February 2 and 6 (MJD=58880 and 58885). 
MIRO is located at Gurushikhar peak in Mount Abu, India, at 
%longitude of 72$^\circ$ 46'47.5'' E, latitude of 24$^\circ$39'08.8'' N, and 
altitude of 1680 m and is operated by Physical Research Laboratory (PRL), Ahmedabad, India. 

% Rozhen
The optical data from National Astronomical Observatory (NAO) Rozhen, Bulgaria were obtained between 2020 January 24 and 26 (MJD=58872 -- 58874). 
We used the 2-m RCC telescope with Andor iKon-L CCD camera (2048x2048 px, 13.5 $\mu$m/pixel) and the 50/70 cm Schmidt telescope with FLI PL-16803 CCD camera (4096x4096 px, 9 $\mu$m/pixel). 
Additional observations were carried out 2020 January 31 to February 2 (MJD=58879 - 58881) at Student Astronomical Observatory (SAO) Plana \citep{ov14} with 35 cm Newton telescope and SBIG STL-11000M CCD Camera (4008x2672 px, 9um/pixel). 
All cameras are equipped with standard photometric UBVRI Johnson-Cousins filters. 	

The data are reduced (including bias subtraction, flat-fielding, and cosmic-ray correction) and analyzed using standard photometry packages from IRAF\footnote{\burl{http://iraf.noao.edu/}}. 
For each image the PSF value is measured and aperture photometry is applied. 
Standard stars from the SDSS DR12 
and VizieR % MIRO
catalogs are used for photometric calibration after applying transformation equations\footnote{\burl{http://www.sdss3.org/dr8/algorithms/sdssUBVRITransform.php##Lupton2005}}. 

%% SIENA photometry

The Astronomical Observatory of the University of Siena observed \src\ in its program devoted to optical photometry of blazars in support of MAGIC. The observatory runs a remotely operated 30 cm Maksutov-Cassegrain telescope installed on a Comec 10micron GM2000-QCI equatorial mount. The focal plane hosts a Sbig STL-6303 camera equipped with a 3072 x 2048 pixels KAF-6303E sensor; Johnson-Cousins BVRI filters are available. Multiple 300\,s images of \src\ were acquired at each visit. After standard dark current subtraction and flat-fielding, images are averaged and aperture photometry is performed on the average frame by means of the MaximDL software package. Reference and check stars in the field of view are selected from the APASS9 \citep{apassDR9} catalog. The reference R magnitudes are derived from those reported in the same APASS9 catalog after conversion between the two different photometric systems, following a formula taken from \cite{munari12}. 

Additionally we use publicly available data in V-band and g-band of ASAS-SN \citep{sh14, ko17}. 

Conversion of magnitudes to energy fluxes is done using zero points of \cite{be98}.  % \komm{Perkins, ASAS-SN, others (?)}.  
Correction for the Galactic extinction is applied using the E(B--V) value of 0.011 from \cite{sf11} 
% \komm{Perkins, ASAS-SN, Rozhen, REM, others (?)} 
and the extinction laws from \citet{cardelli89}.

We have performed observations of optical spectra of the quasar \src\ using the 4.3~m Lowell Discovery Telescope (LDT; Lowell Obs., Flagstaff, AZ) equipped with the DeVeny spectrograph and the Large Monolithic Imager (LMI), in response to the detection of the source at VHE by MAGIC on 2020 January 21 (MJD=58869). 
We employed a grating setting of 300 grooves per millimeter, which provides spectra from 3300\AA\ to 7500\AA\ with a dispersion of 2.17 \AA\ per pixel, a blaze wavelength of 5000\AA\, and a slit width of 2.5$^{\prime\,\prime}$. The spectroscopic observations were performed on 2020 January 28 (MJD 58876.578), February 8 (MJD 58887.497) and 25 (MJD 58904.368). During this month the brightness of the quasar fell from 14.6\,mag to 16.25\,mag in R band. Each observation of the quasar consisted of 3 exposures of 600\,s (900\,s on February 25). Observations of a comparison star HD126944 of A type, located $\sim$86$^{\prime}$ from the quasar, were performed before and after target observations, with two 30~s exposures for each observation. Bias and flat-field images were obtained regularly. The LDT allows a switch between different instruments in 2-3 min. Therefore, photometry of the quasar using the LMI in V and R filters were performed just before or after spectral observations to calibrate the spectra. The observations are reduced using programs written in IDL (v.8.6.1) that implement the technique described in \cite{Vacca2002} developed for reduction of spectra obtained with SpeX at the NASA Infrared Telescope Facility on Mauna Kea.

\subsection{NIR}

The NIR observations were carried out with the camera CANICA \citep{ca17}, along with the Guillermo Haro 2.1m telescope (OAGH), located at Cananea Sonora Mexico. 
The camera is based on a Hawaii detector of 1024 by 1024 pixels with a plate scale of 0.32 arcsec per pixel. 
The data are part of the monitoring program "NIR photometry of AGNs with Gamma Ray emission detected by \fermi" . 
Relative photometry is obtained with respect to the 2MASS point sources included in the field of view (5.5 arcmin). 
Absolute fluxes are obtained adopting the zero point values of 2MASS derived by \cite{cwm03}.
The host galaxy is not detected in IR by the 2MASS survey in any of the three bands, resulting in an upper limit of H $\sim$ 17.7 mag, much weaker than 13--11 mag observed by CANICA during the investigated period. 
Therefore the effect of the host galaxy is negligible.  

\subsection{Radio}

% Metsahovi
The 37 GHz observations were made with the 13.7 m diameter Mets\"ahovi radio telescope. 
The detection limit of the telescope at 37 GHz is on the order of 0.2 Jy under optimal conditions. 
Data points with a signal-to-noise ratio < 4 are handled as non-detections. 
The flux density scale is set by observations of DR~21. 
Sources NGC 7027, 3C~274 and 3C~84 are used as secondary calibrators. 
A detailed description of the data reduction and analysis is given in \cite{te98}. 
The error estimate in the flux density includes the contribution from the
measurement rms and the uncertainty of the absolute calibration.

We also use publicly available 15\,GHz OVRO data \citep{ri11} and 8.63\,GHz data from Badary RT32 reported in \citep{kh20}.

We requested Director’s Discretionary Time (DDT) with the Very Long Baseline Array (VLBA) following the gamma-ray activity and VHE detection of the quasar \src\, and were granted 6 epochs of observations of the source separated by approximately 1 month intervals (ID~BD227), with 8 hrs per epoch. Thus far, we have obtained 3 epochs of observations under the project, on 2020 March 8, May 10, and June 6 (MJD=58916, 58979, 59006). %One more epoch is observed on July 4
The observations were performed with all 10 antennas in continuum, dual circular polarization mode at 43\,GHz using 4 intermediate frequency bands (IFs), each of 64\,MHz width. The data were correlated at the National Radio Astronomy Observatory (NRAO, Soccoro, NM) using the VLBA DiFX software correlator. Five sources were observed at each epoch (\src\, 3C~279, 3C~345, PKS 1055+18, and B2 1308+326), with 60\% of the time devoted to the main target, \src. The sources 3C~279, 3C~345, and PKS 1055+018 are used for fringe finding during the correlation. The data are reduced using the Astronomical Image Processing System software \citep[AIPS,][]{MKG1996}) and Difmap package \citep{Difmap} in the same manner as described in \cite{Jorstad2017}, except without averaging of the final calibrated data over IFs. We use the sources observed along with \src\ to perform absolute calibration of the electric vector position angle (EVPA), since these sources are monitored in the VLBA-BU-BLAZAR program\footnote{\burl{www.bu.edu/blazars/VLBAproject.html}}, so that their polarization properties at 43~GHz are known. 

\section{Results}  \label{sec:res}
\begin{figure*}%[tp]
  \centering
  \includegraphics[width=0.9\textwidth, height=0.9\textheight]{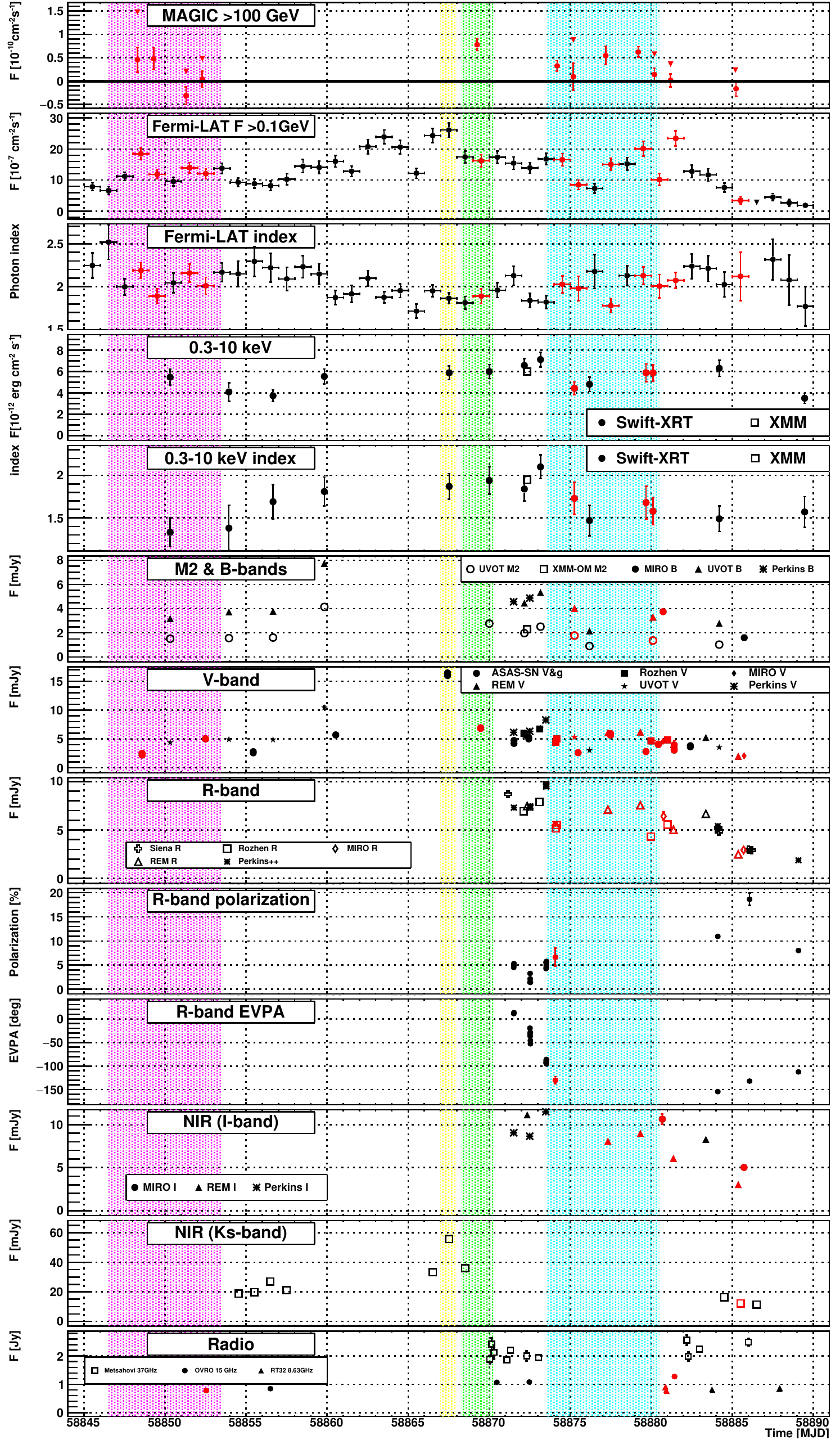}
  \caption{
     MWL light curve of \src\ between 2019 December 28 (MJD=58845) and 2020 February 11 (MJD=58890); see titles and legends of individual panels. 
     Optical and UV observations are corrected for the Galactic attenuation. 
     The points in red are contemporaneous ($\pm12$\,hr) with MAGIC observations. 
     The shaded regions show the time ranges of the four considered flare evolution periods (see Table~\ref{tab:periods}).
     Flux upper limits in the first two panels are shown  with downward triangles.
     \label{fig:mwl_lc}
}
\end{figure*}

\begin{figure}%[tp]
  \centering
  \includegraphics[width=0.45\textwidth]{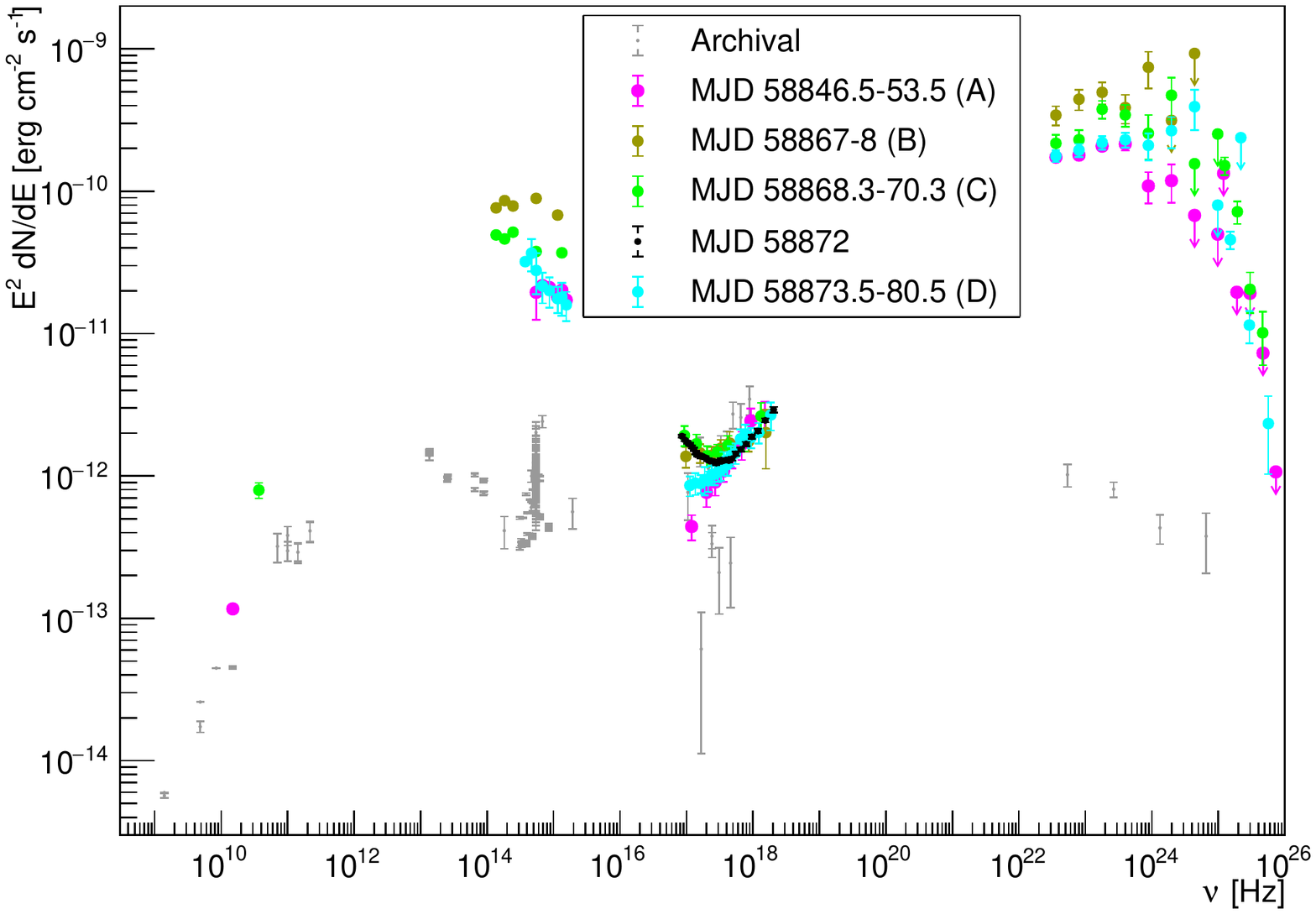}
  \caption{
     MWL SED of \src\ in 2020 January/February. 
Points follow the colors of the shaded regions in Fig.~\ref{fig:mwl_lc}.
In addition in black XMM points taken on 2020 January 24 (MJD=58872) are shown. 
MAGIC points are not corrected for the EBL absorption. 
Gray points show archival data (most from the ASI Space Science Data Center, but also including low-state \fermi{} spectrum and lowest and highest X-ray spectrum from \swift).
     \label{fig:mwl_sed}
}
\end{figure}

In Fig.~\ref{fig:mwl_lc} we present the MWL light curve summarizing the evolution of the flare. 
Based on the VHE state of the source we define three periods selected for the further spectral analysis:
A: 2019 December 29 to 2020 January 5 (MJD=58846.5 -- 58853.5): without VHE gamma-ray detection, 
C: 2020 January 20 to 22 (MJD=58868.3 -- 58870.3): VHE gamma-ray flare, 
D: 2020 January 25 to February 1 (MJD=58873.5 - 58880.5): detection over longer time scale. 
In addition we define the fourth period:
B: 2020 January 19 to 20 (MJD=58867 -- 58868), which does not have simultaneous MAGIC data, but contains the peak of the optical and IR flare as well as one of the local peaks of HE emission. 
The four periods (referred to throughout the paper as periods A-D) are summarized in Table~\ref{tab:periods}. 
In each period we construct a broadband SED (see Fig~\ref{fig:mwl_sed}).

\begin{table}[]
    \centering
    \begin{tabular}{c|c|c}
         Period & MJD & comment \\\hline
         \textcolor{magenta}{A} & 58846.5 - 58853.5 & pre-flare \\
         \textcolor{olive}  {B} & 58867 - 58868 & optical flare \\
         \textcolor{green}  {C} & 58868.3 - 58870.3 & VHE flare \\ % 0.26
         \textcolor{cyan}   {D} & 58873.5 - 58880.5 & post-flare \\
    \end{tabular}
    \caption{Periods selected for the SED modeling (the colors of the period tag correspond to colors used in the figures)}
    \label{tab:periods}
\end{table}

In the case of gamma-ray data all observations performed within a given time window are  summed. 
X-ray data available from different observations are stacked for the period D, while in the other periods a single {\em Swift}-XRT observation is available and used. 
On the other hand, the low-energy data (radio up to UV) have mostly lower uncertainties and hence are more sensitive to variability. 
Therefore, if more than one measurement was taken at a given time period we average all the measurements and take the standard deviation of the measurements as the measure of its uncertainty. 
A similar approach for constructing a broadband SED has been applied, e.g., in \citet{acc18}.

\subsection{VHE gamma-ray emission}
\begin{figure}%[t!]
  \centering
  \includegraphics[width=0.48\textwidth]{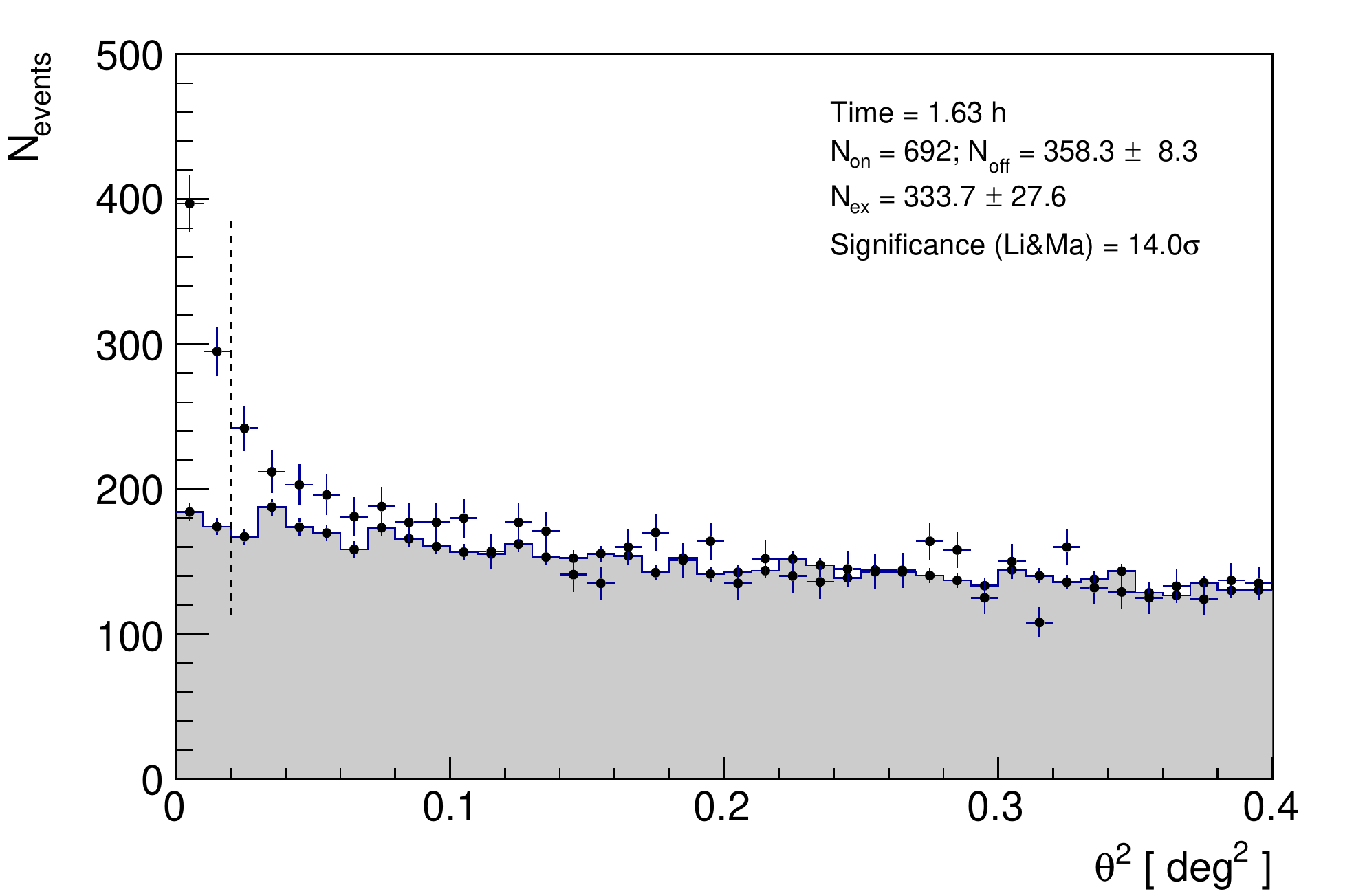}
  \caption{
    Distribution of the squared angular distance between the source nominal and event reconstructed directions (points) and corresponding background estimation (shaded region) for the night of 2020 January 20 (MJD=58868) of MAGIC observations. Vertical line shows the cut value below which the event statistics are given in the inset. 
 }
  \label{fig:magic_sign}
\end{figure}

The first detection of the VHE gamma-ray emission from \src\ was achieved on 2020 January 20 (MJD=58868). 
A highly significant detection of $14.3\sigma$ is obtained in 1.6\,hr of effective time (see Fig.~\ref{fig:magic_sign}).
In the subsequent period of 2020 January 26 to February 1 (MJD=58874 -- 58880) further hints of signal are obtained, with the highest significance of the excess ($6.6\sigma$) on the night of 2020 January 31 (MJD=58879), which has also the longest exposure of 2.5\,hr. 

During the flare (period C), the flux observed by MAGIC above $100\,$GeV reached $(7.8 \pm 1.3_{\rm stat}) \times 10^{-11} \mathrm{cm^{-2} s^{-1}}$.
The observed spectrum in this period can be described by a power-law:
$dN/dE = (2.49 \pm 0.31_{\rm stat}) \times 10^{-9} (E/100\,\mathrm{GeV})^{-4.22\pm 0.24_{\rm stat}}\mathrm{[TeV^{-1} cm^{-2} s^{-1} ]}$. 
Correcting for the EBL absorption according to \cite{do11}, the unattenuated   spectrum can be described as
$dN/dE\,e^{\tau_{\rm EBL}(E)}= (4.04 \pm 0.54_{\rm stat}) \times 10^{-9} (E/100\,\mathrm{GeV})^{-3.57\pm 0.29_{\rm stat}}\mathrm{[TeV^{-1} cm^{-2} s^{-1} ]}$.

After the flare (period D), significant gamma-ray emission is detected again with $5.6\sigma$, but at about half the flare level: $F_D(>100\,\mathrm{GeV}) = (3.9 \pm 0.6_{\rm stat}) \times 10^{-11} \mathrm{cm^{-2} s^{-1}}$.
The observed spectrum in this period can be described:
$dN/dE = (0.91 \pm 0.13_{\rm stat}) \times 10^{-9} (E/100\,\mathrm{GeV})^{-3.90\pm 0.25_{\rm stat}}\mathrm{[TeV^{-1} cm^{-2} s^{-1} ]}$. 
Correcting for the EBL absorption according to \cite{do11}, the spectrum can be described as
$dN/dE\,e^{\tau_{\rm EBL}(E)} = (1.64 \pm 0.22_{\rm stat}) \times 10^{-9} (E/100\,\mathrm{GeV})^{-2.87\pm 0.36_{\rm stat}}\mathrm{[TeV^{-1} cm^{-2} s^{-1} ]}$. 
Despite enhancement of the VHE gamma-ray flux by a factor of two, the spectral indices during and after the flare are consistent within $1.5\sigma$. 
It should be noted however that, in particular in period D, the uncertainties of the photon index are large. 

\begin{figure}%[t!]
  \centering
  \includegraphics[width=0.48\textwidth]{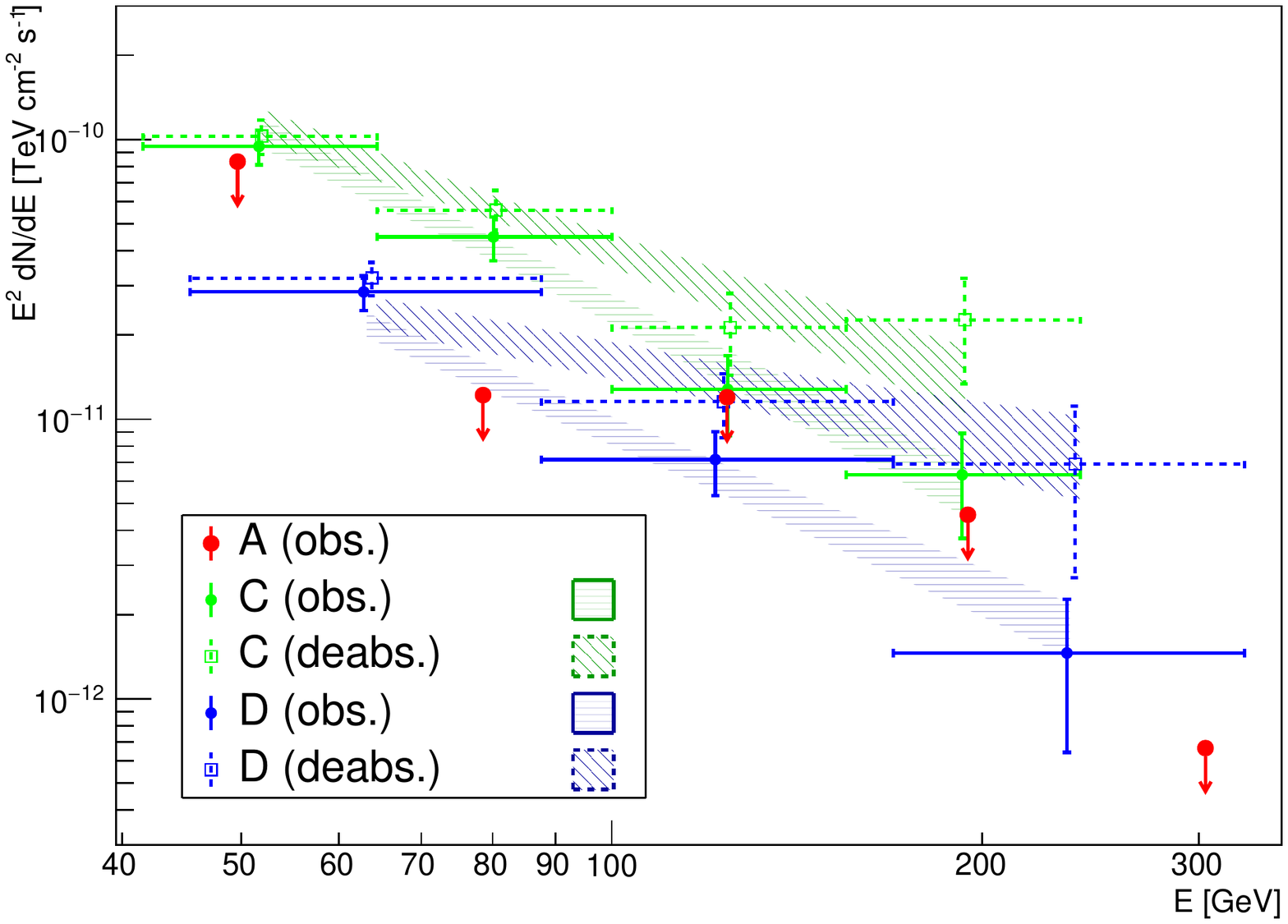}
  \caption{
    SED of \src\ observed by MAGIC in periods A, C and D (see legend): observed (filled circles) and corrected for EBL absorption (empty circles). 
 }
  \label{fig:magic_sed}
\end{figure}

Before the flare (period A), possibly due to shorter observations under less favorable zenith angle, no significant emission is detected and only a 95\% C.L. limit of $<4.1 \times 10^{-11} \mathrm{cm^{-2} s^{-1}}$ can be placed for the flux above 100\,GeV. 
The limit is comparable to the emission detected from the source in period D. 
The SED of \src\ observed by MAGIC in different periods is shown in Fig.~\ref{fig:magic_sed}. 

\subsection{HE gamma rays}
The first detection of a HE outburst from B2\,1420+326 was reported in 2018 December \citep{cip18}, with a flux increase by more than two orders of magnitude with respect to the average 4FGL value and significant spectral hardening. A similar spectral hardening has been reported in 2019 July \citep{ang19}, where the first evidence of $>10$\,GeV photons was provided, and again in 2020 January \citep{cc20}.

The daily \textit{Fermi}-LAT light curve is shown in Fig.\,\ref{fig:mwl_lc}, including the flux (second panel) and photon index (third panel). Both the flux and the photon index are significantly variable in this time interval, based on a simple $\chi^2$ test. The \textit{Fermi}-LAT recorded a peak daily flux (E $>$ 100 MeV) of $(2.6\pm0.2)\times10^{-6}$\,cm$^{-2}$\,s$^{-1}$ on 2020 January 19 (MJD=58867), 
corresponding to about $300$ times the average value reported in the 4FGL catalog. The highest-energy photon observed by the \textit{Fermi}-LAT was recorded two days prior (2020 January 17, MJD=58865), with an energy of $\sim150$\,GeV, providing the first indication of VHE emission from \src \footnote{The probability of this photon being associated with the target source is $>99.9$\%, as obtained from the \texttt{gtsrcprob} tool.}. Accordingly, the \textit{Fermi}-LAT recorded the hardest daily spectrum on the same date, with a photon index of 1.72 $\pm$ 0.08, a flux of $(1.2\pm0.2)\times10^{-6}$\,cm$^{-2}$\,s$^{-1}$ and a TS of 496. We also note that the photon index was consistently harder than the 4FGL catalog value ($2.38\pm0.07$) during most of the time range shown in Fig\,\ref{fig:mwl_lc}.

\begin{figure}%[t!]
  \centering
  \includegraphics[width=0.48\textwidth]{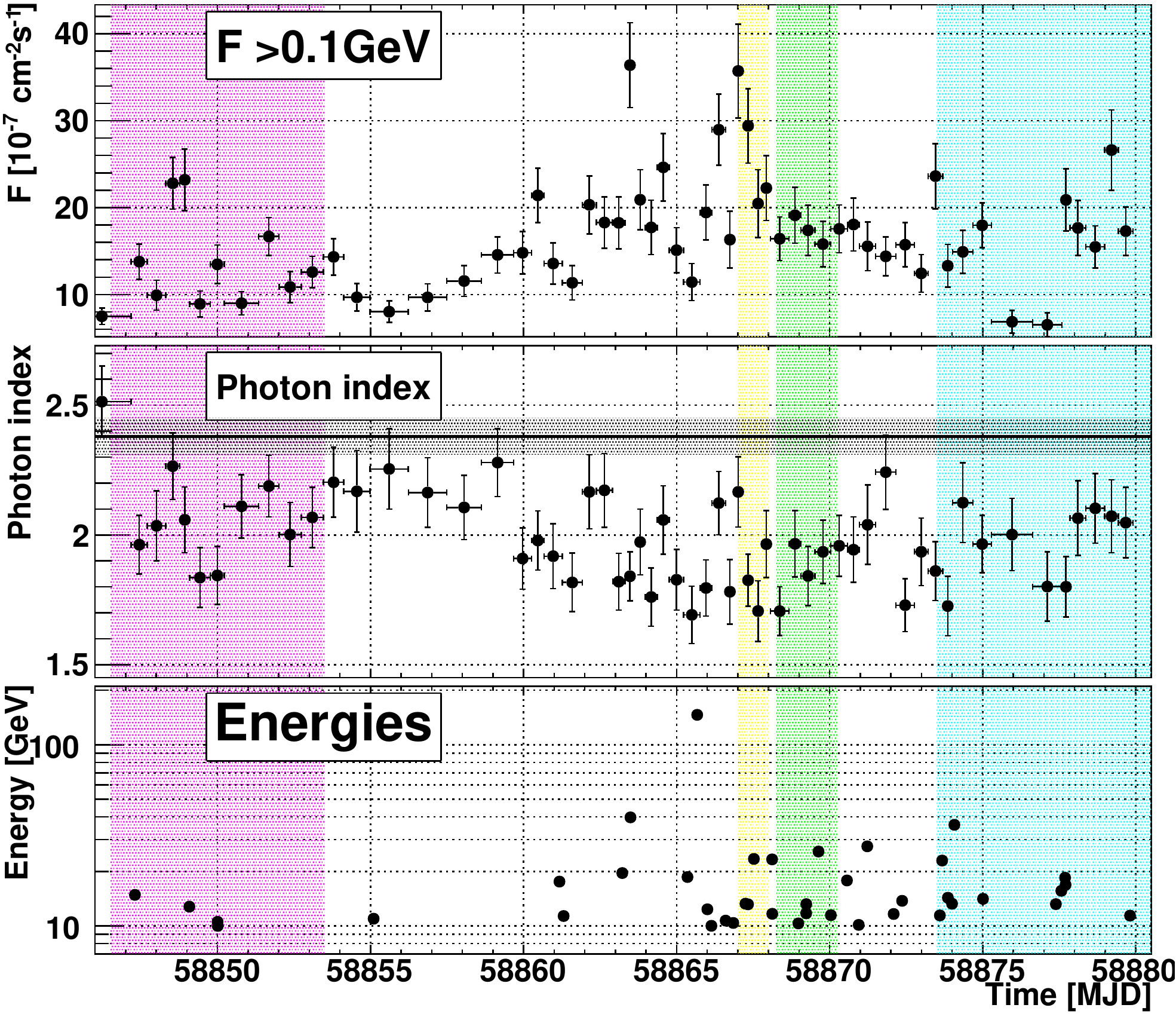}
    \caption{
    \textit{Fermi}-LAT adaptively binned light curve of \src\ in January 2020. The vertical shaded areas follow the color coding of Fig.\,\ref{fig:mwl_lc}. \textit{Top panel}: 0.1-300 GeV photon flux. \textit{Middle panel}: power-law photon index (the horizontal shaded area represents the average value from the 4FGL with its uncertainty). \textit{Bottom panel}: E $>$ 10\,GeV photons associated with \src\ with probability $>80\%$.
 }
  \label{fig:lat_ab}
\end{figure}

The adaptively binned light curve is shown in Fig.\,\ref{fig:lat_ab}. 
The shortest adaptive bin is centered on 2020 January 19, 00:17:45 (MJD=58867.012)%00:17:44.794 UTC 
, and has a width of $\sim6$ hours. The highest flux is recorded in the same bin, reaching a value of $(3.6\pm0.5)\times10^{-6}$\,cm$^{-2}$\,s$^{-1}$, i.e. 400 times higher than the 4FGL value. 

We note that a test for spectral curvature was performed in all time bins of all light curves, and a power-law spectrum was found to be the best representation in all time intervals. 

Additionally, we perform a likelihood fit over the full time interval included in Fig.\,\ref{fig:mwl_lc}, to characterize the average source properties in the HE band in this flaring state. 
For this time interval, the LogParabola model is preferred ($TS_{curv}=36$ 
\footnote{$TS_{curv} = 2(\ln{L_{LP}} - \ln{L_{PL}})$ is used to check if a statistically significant curvature is detected using a LogParabola model compared with the PowerLaw model, where $\ln{L}$ indicates the logarithm of the likelihood for each model. A source is considered to have a statistically significant curvature if $TS_{curv}>25$.})
with respect to a simple power law to describe the gamma-ray spectrum of the source.
It yields a photon flux $(1.09\pm0.02)\times10^{-6}$\,cm$^{-2}$\,s$^{-1}$, and spectral parameters $\alpha=1.97\pm0.02$, and $\beta=0.07\pm0.01$~\footnote{The functional form of the LogParabola spectral model is $dN/dE=N_0(E/E_b)^{-[\alpha+\beta\log(E/E_b)]}$, where $N_0$ is the normalization, $E_b$ is the reference energy at which $N_0$ is measured, $\alpha$ is the slope and $\beta$ is the curvature parameter.}. We also verify that there are no new significant point sources in addition to the initial 4FGL model during this period.

Finally, as mentioned in Section\,\ref{sec:obs}, we perform separate likelihood fits corresponding to the periods used to build time-resolved SEDs. The results of these fits are summarized in Table\,\ref{tab:lat_seds}, together with the ones for the quiescent state.

%%%%%%%%%%%%%%%%%%%%%%%%%%%%%%%%%%%%%%%%%%%%%%%%%%%%%%%%%%%%%%%%%%%%%%%%%
\begin{table}[htbp]
    \begin{center}
    \begin{tabular}{cccc}
    \hline
    \hline
    State & Flux$^\mathrm{a}$ & Photon index & TS\\
    \hline
    Quiescent & $0.86\pm0.13$ & $2.40\pm0.08$ & 218\\
    A & $115\pm6$ & $2.04\pm0.04$ & 1942\\
    B & $260\pm20$ & $1.88\pm0.07$ & 688\\
    C & $160\pm20$ & $1.88\pm0.06$ & 927\\
    D & $121\pm7$ & $1.93\pm0.04$ & 2280\\
    \hline
    \hline
    \end{tabular}
    \end{center}
    $^\mathrm{a}$ Total flux in the energy range 0.1-300 GeV in units of $10^{-8}$\,cm$^{-2}$\,s$^{-1}$.
\\
\caption{Results of \textit{Fermi}-LAT analysis on the different activity states of \src. The time periods are defined in Table\,\ref{tab:periods}.}\label{tab:lat_seds}

\end{table}
%%%%%%%%%%%%%%%%%%%%%%%%%%%%%%%%%%%%%%%%%%%%%%%%%%%%%%%%%%%%%%%%%%%%%%%%%

%
\subsection{X-ray}
There is no strong variability of the X-ray flux in the investigated period with an increasing trend from 2020 January 05 (MJD=58853) to 2020 January 25 (MJD=58873), where a peak flux a factor of 2 higher is observed.
However there is a clear variability of the X-ray spectral index from hard values before the flare to much softer at the time around the optical and gamma-ray flares, back to hard values. 
This indicates a shift of the synchrotron peak of the source and connected with it, the shift of the crossing point between the synchrotron and the IC component. 
This is clearly visible in the X-ray spectra during the flare and during {\em XMM-Newton} observations (see Fig.~\ref{fig:mwl_sed}). 
Results of the spectral fits to individual days of \swift{} observations are given in Appendix~\ref{sec:swift}.

\subsection{Optical}
Compared to the historical measurements, the optical emission is $\sim1.5$ orders of magnitude higher throughout the investigated period. 
Moreover, during that period a strong optical flare is observed on 2020 January 19 (MJD=58867) with a variability time scale of the order of a few days.
V-band observations performed during one of the \fermi{} peaks show an increase by nearly an order of magnitude with respect to observations at the beginning of Period A. 
Similar variability pattern is seen also in IR and UV ranges.  
However, the spectral shape in the IR-UV range varies during the investigated period, with the spectrum becoming bluer (harder) during the optical flare (see Fig.~\ref{fig:mwl_sed}). 
This is consistent with the X-ray behavior of the source that also suggests a shift of the synchrotron peak position. 
Recently other occurrences of comparable flaring activity in the optical had been observed \citep[e.g., in July 2019,][]{marchini19}, reporting a $R=13.7$ magnitude even slightly brighter than the brightest $R$ point in Fig \ref{fig:mwl_lc}. 

Also optical polarization shows interesting behavior with a dip of the polarization percentage at a few per cent level and a concurrent rotation by $\sim150^\circ$.
Similar EVPA rotation during low level of polarization has been also seen contemporaneous with VHE emission in, e.g., PKS\,1510--089 \citep{al14}. 
Large variability of EVPA (down to a time scale of 3 hours) has been also seen in PKS\,0736$+$017, another FSRQ during the period of VHE gamma-ray detection \citep{ab20}.
% TON 599 - rotation by 40 deg and P=10-20%
% PKS1441 (VERITAS paper) - rotation by 25deg (but large P)

%

\subsection{Optical Spectroscopy}\label{sec:spectro}

\begin{figure}%[t!]
  \centering
  \includegraphics[width=0.48\textwidth]{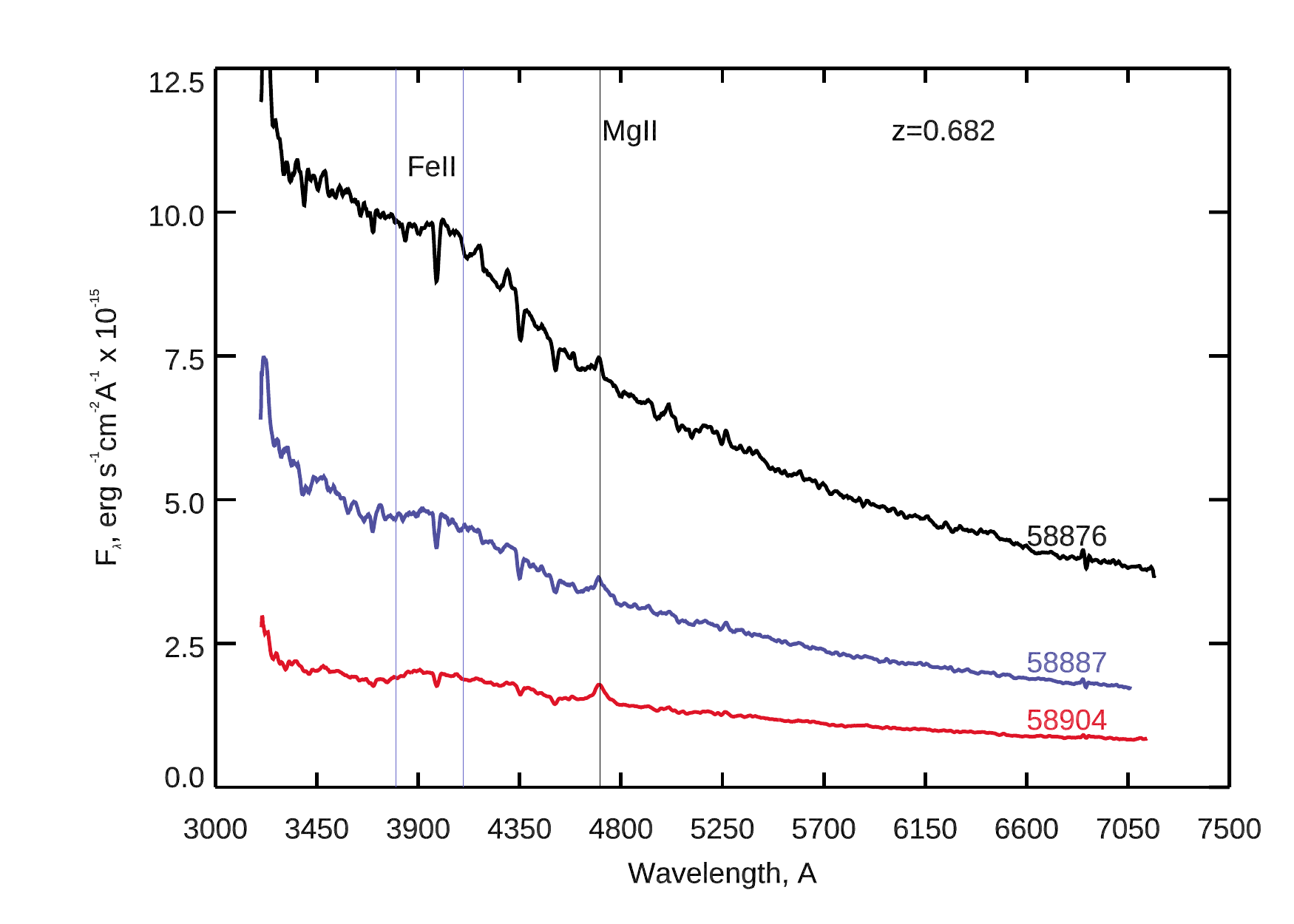}
  \caption{
    Optical spectra of \src\ obtained with the LDT; the spectra are in the observer's frame.
 }
  \label{fig:LDTspectra}
\end{figure}

Fig.~\ref{fig:LDTspectra} presents the optical spectra of \src\ obtained with LDT at three epochs 
of different activity states (taken mostly after the gamma-ray flaring activity). The spectra show the presence of MgII emission line at $\lambda=$4706 \AA\ (rest wavelength of 2798 \AA) and  
a bump between 3800 \AA\ and 4100 \AA\ (rest frame $\sim$2260$-$2450 \AA). The latter
appears to be part of an FeII emission complex, whose strongest UV lines fall in the 2200-2600 \AA\ range \citep[e.g., ][]{Baldwin2004}. The spectra also include three prominent absorption lines at $\lambda\sim$4001\AA\, 4337\AA\, and 4477\AA\ that intensify as the quasar brightens. The absorption lines are most likely intervening MgII systems at redshifts of $z\approx0.43$, 0.55, and 0.60, respectively. 

\begin{figure}%[t!]
  \centering
  \includegraphics[width=0.48\textwidth]{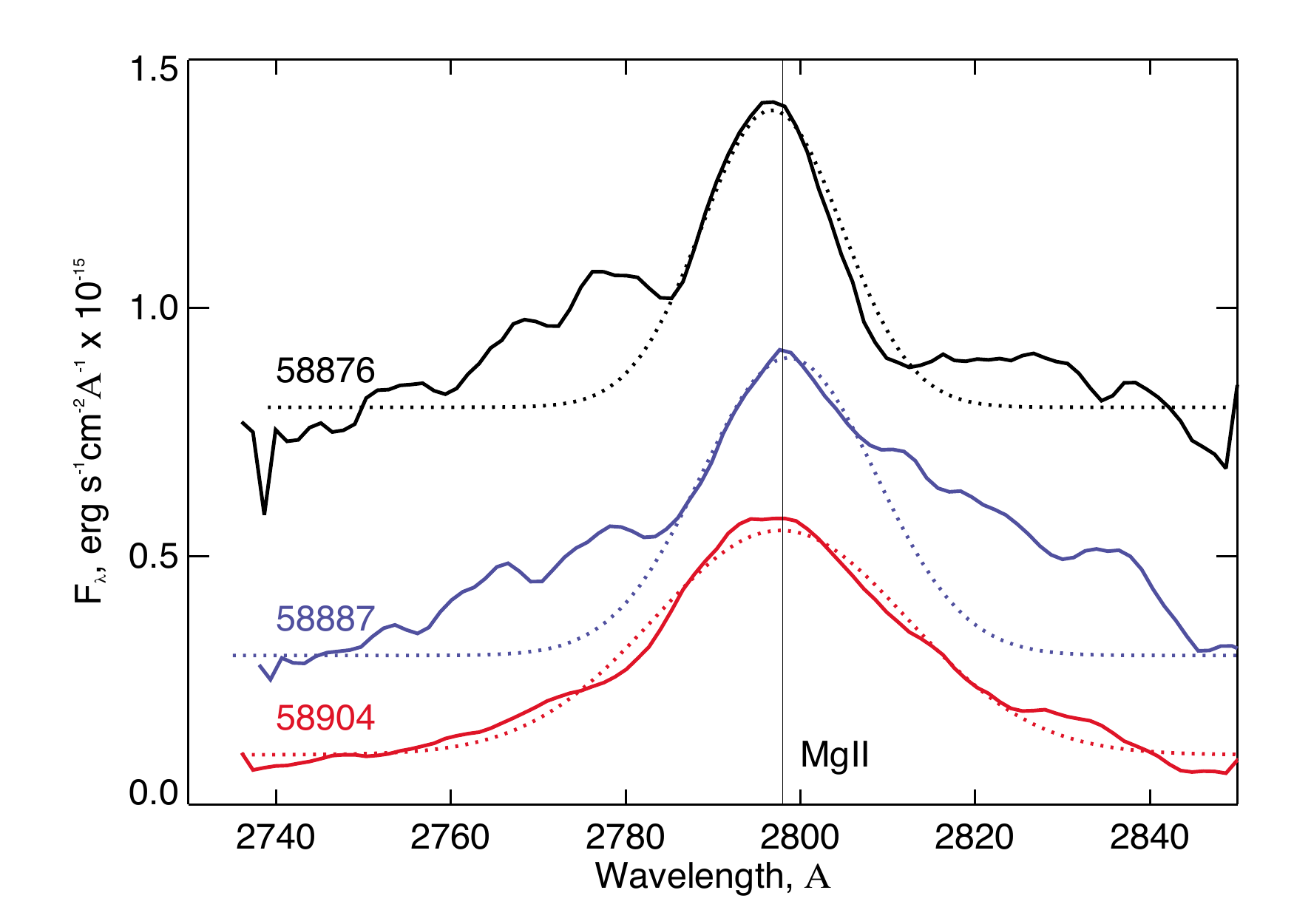}
  \caption{
    Observed (solid) and modeled (dotted) MgII line of \src\ in the rest frame at three epochs. The lines on 2020 January 28 (MJD=58876), February 8 (MJD=58887), and February 25 (MJD=58904) are shifted by values of 0.8 0.3, and 0.1, respectively, for clarity.
 }
  \label{fig:MgIIfit}
\end{figure}

\begin{figure}%[t!]
  \centering
  \includegraphics[width=0.48\textwidth]{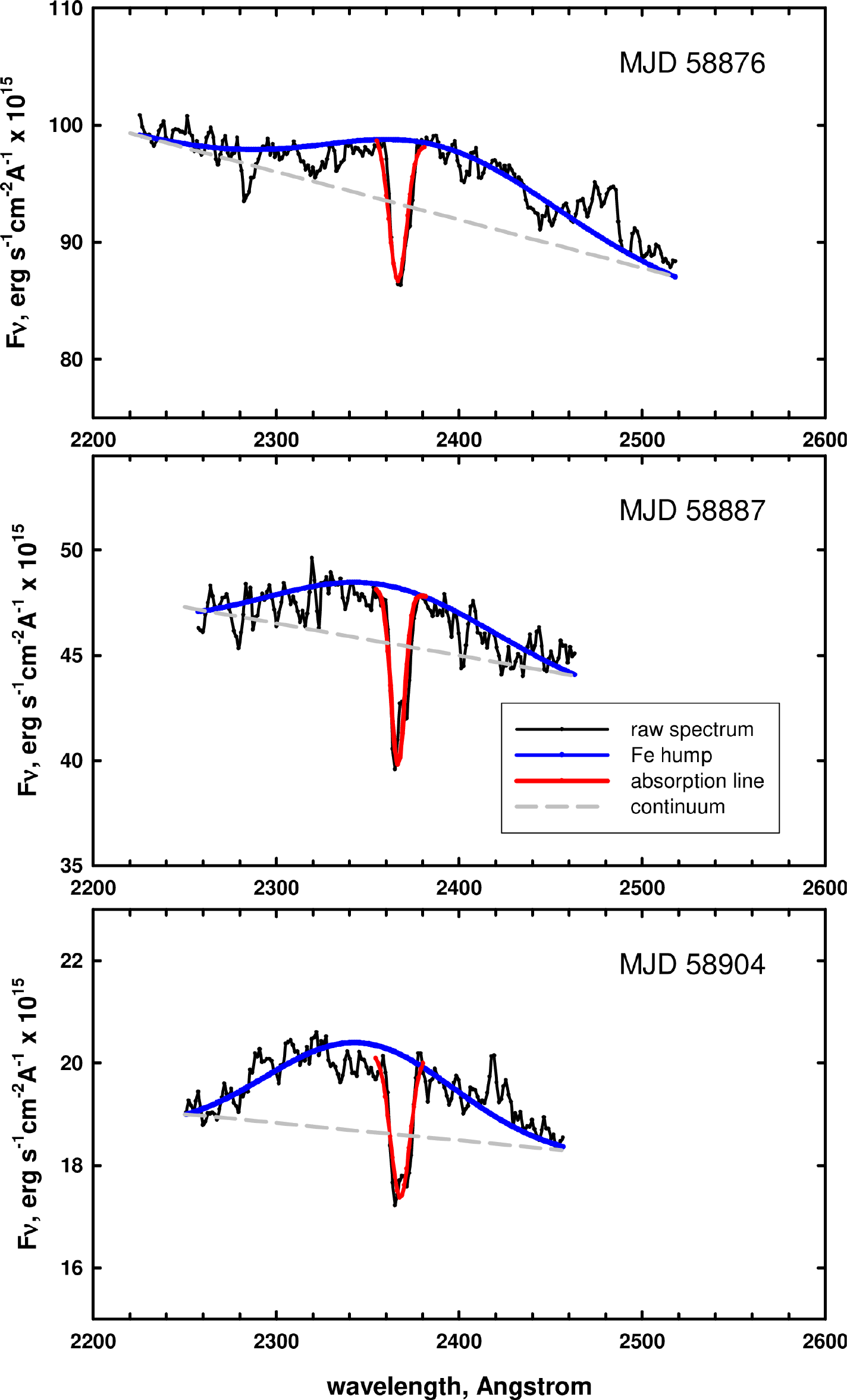}
  \caption{
    Observed (black) and modeled (blue) FeII bump of \src\ at three epochs in the rest frame; red lines represent Gaussian fits to absorption lines, and gray dashed lines indicate the continuum level.
 }
\label{fig:FeIIfit}
\end{figure}

\begin{figure}%[t!]
  \centering
  \includegraphics[width=0.48\textwidth]{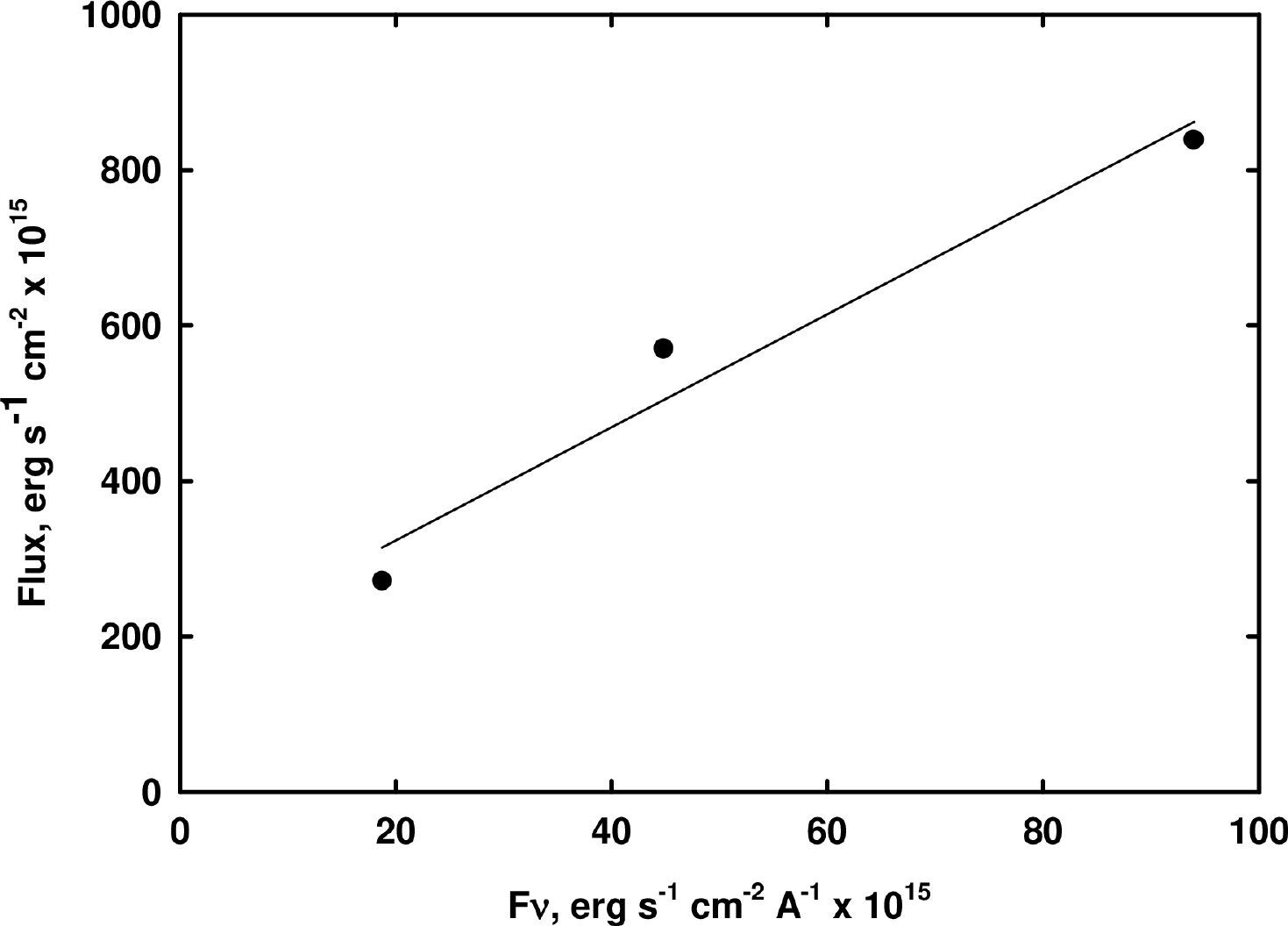}
  \caption{
    Dependence between flux of the FeII bump and flux density of the continuum
    at different epochs (points show the measurements, and the line is a linear fit). 
 }
\label{fig:Feregress}
\end{figure}

We analyze the characteristics of the MgII emission line and FeII bump as a function of continuum brightness. Fig.~\ref{fig:MgIIfit} shows an approximate Gaussian fit to the MgII line, while Fig.~\ref{fig:FeIIfit} plots similar modeling of the FeII bump at all three epochs. 
The parameters of the Gaussian and velocities of clouds, as well as the flux of the lines, are given in Table~\ref{tab:LineParm}, although we are unable to estimate the velocity of gas producing FeII lines, since the bump consists of $>$100 lines.
There is a significant difference between the MgII line and FeII bump behavior: i) the flux of the MgII line remains constant within 1$\sigma$ uncertainty independent of the continuum brightness, while the FeII bump increases in flux with the continuum level (see Fig.~\ref{fig:Feregress}); ii) the central wavelength of the MgII line fits does not show a shift with respect to the rest wavelength, while $\lambda_\circ$ of the FeII bump 
shifts toward the blue side as the time after the VHE event passes. Unfortunately, it is not possible to distinguish whether the shift is due to a relative change of the brightness of lines that form the FeII bump, or due to gas motion toward the observer; and iii) the FWHM of the FeII bump is very stable despite the correlation of its flux with the continuum level, while the velocity of gas where the MgII line originates increases with time after the VHE event. 
We note also a significant change of the equivalent width (EW) of the MgII line with the continuum, with EW decreasing as the continuum rises. 
This questions the identification of \src\ as a FSRQ, however, Table~\ref{tab:LineParm} shows that at the lower levels of activity EW$>$5\AA\ for the MgII line\footnote{EW of 5\AA\ is the classical threshold between BL Lac and FSRQ objects, see, e.g., \cite{sa96}}.

\begin{table*}[]
    \centering
    \begin{tabular}{r|r|r|r|r|r|r|r|r}
    \hline
MJD & Line &$\lambda_\circ$& Amp & FWHM & v & Flux & S$_{cont}$& EW \\
(1) & (2) & (3) & (4) & (5) & (6) & (7) & (8) & (9)\\
\hline 
58876.578&MgII&2795$\pm$2&0.37$\pm$0.04&18.4$\pm$1.3&1982$\pm$140&13$\pm$3&7.1$\pm$0.2&2$\pm$1 \\
58887.497&MgII&2798$\pm$3&0.36$\pm$0.06&26.0$\pm$2.6&2787$\pm$278&18$\pm$4&3.3$\pm$0.2&6$\pm$3 \\
58904.368&MgII&2797$\pm$2&0.31$\pm$0.02&28.6$\pm$0.7&3074$\pm$75&16$\pm$2&1.5$\pm$0.2&11$\pm$2 \\
58876.578&FeII&2390$\pm$5&0.49$\pm$0.05&91$\pm$3& \nodata &84$\pm$9&9.2$\pm$0.5&\nodataa \\
58887.497&FeII&2357$\pm$5&0.23$\pm$0.06&91$\pm$2& \nodata &58$\pm$5&4.5$\pm$0.2&\nodataa \\
58904.368&FeII&2348$\pm$5&0.15$\pm$0.03&91$\pm$1& \nodata &27$\pm$7&1.8$\pm$0.3&\nodataa \\
\hline
\end{tabular} 
\caption{Parameters of Lines. Columns as follows: (1) epoch of observations; (2) emission line; (3) central wavelength of Gaussian fit in \AA\; (4) amplitude of Gaussian fit in units of 10$^{-15}$erg~cm$^{-2}$s$^{-1}$\AA$^{-1}$; (5) full width of Gaussian at half-maximum (FWHM) in \AA\; (6) velocity of gas in km~s$^{-1}$; (7)  flux of line in 10$^{-15}$erg~cm$^{-2}$s$^{-1}$; (8) flux density of the continuum at the peak of line in 10$^{-15}$erg~cm$^{-2}$s$^{-1}$\AA $^{-1}$; (9) equivalent width in \AA.}
\label{tab:LineParm}
\end{table*}  

The increase in the flux of the FeII bump with the continuum and a possible motion of gas producing FeII lines toward the observer are quite interesting. The short time lag between the continuum and line variability suggests that the FeII emission-line region is
much smaller than the region producing the MgII line and/or lies close to the line of
sight. It is possible that the non-thermal jet is interacting with an FeII-emitting cloud, while the MgII emission is excited by the underlying thermal accretion disk continuum, which varies on a much longer time scale.

We adopt an approach suggested by \cite{Ghisellini2014} (see also references therein),
who  used an estimate of the accretion disk (AD) luminosity based on the luminosity of the BLR, L$_{AD}\approx$10 L$_{BLR}$.  
The known flux density of the MgII line, combined with the BLR template constructed by \cite{VandenBerk2001} for a composite emission spectrum of a quasar using SDSS spectra, allow us to estimate the total luminosity of the BLR in \src\, $L_{BLR}$ = (1.8$\pm$0.2)$\times$10$^{45}$\,erg~s$^{-1}$. This, for luminosity distance of 4256.4~Mpc% 4125~Mpc, 
 translates to $L_{AD}\approx2\times10^{46}$\,erg~s$^{-1}$, with an uncertainty of a factor of $\sim 2$, as suggested by \cite{Ghisellini2014}. 
Interestingly, the obtained luminosity of the accretion disk is rather high, in the high-end part of values shown for other sources \cite{Ghisellini2014}.

\subsection{Radio}\label{sec:vlbi}
Moderate variability is seen in radio observations (see also Appendix~\ref{app:longterm}). 
OVRO data during the investigated period show a gradual increase of the flux.  
No monotonous behavior is seen in flux at 37 GHz by Mets\"ahovi, but a constant fit can be rejected at $\chi^2 / N_{\rm dof}=51/21$. 
The amplitude of the variability is $\sim 10\%$.

\begin{figure}%[tp]
  \centering
\includegraphics[width=0.49\textwidth]{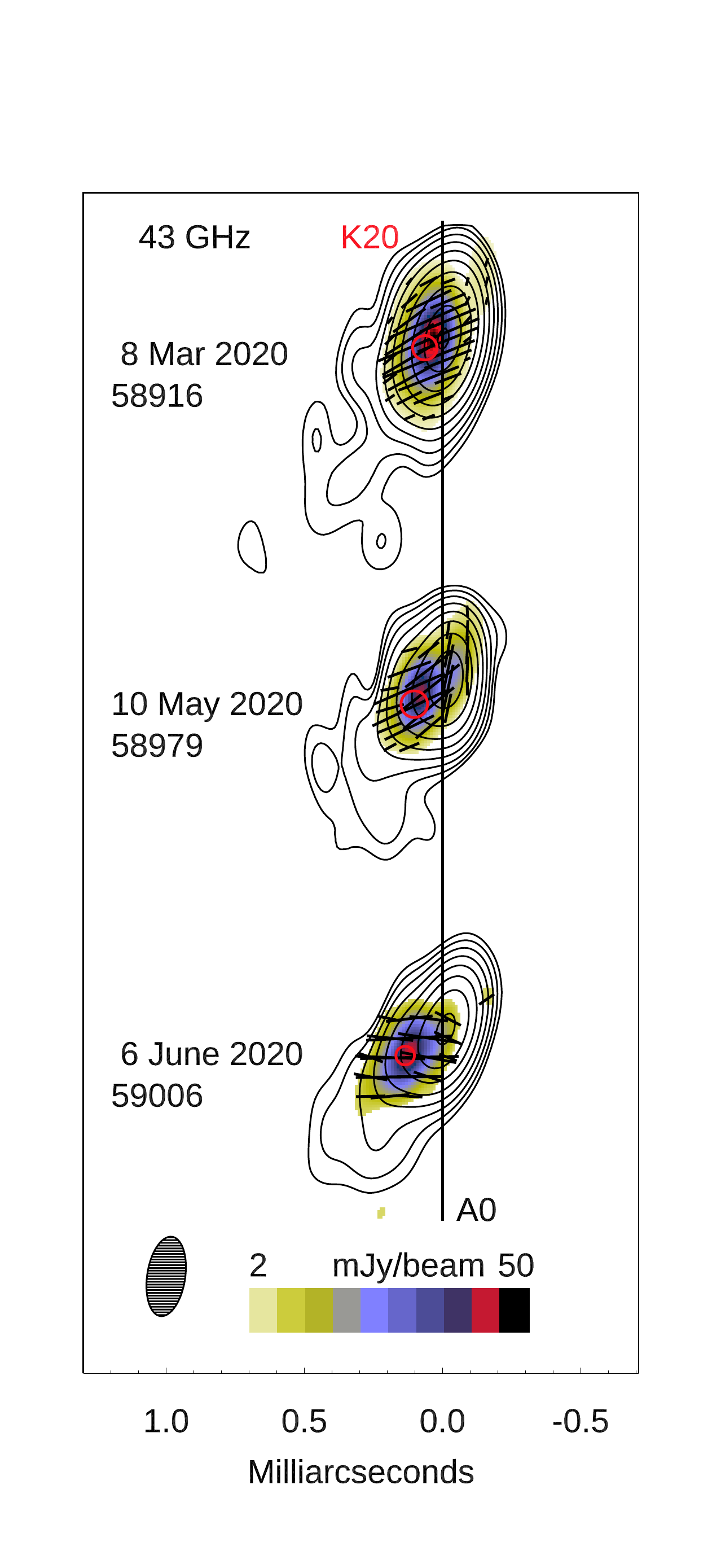}  
  \caption{Total (contours) and polarized (color scale) intensity maps of \src\  obtained with the VLBA at 43~GHz; the peak total intensity is 1.14 Jy/beam; the beam is displayed in the bottom left corner; the contours are 0.25, 0.5, 1,...64, 90\% of the peak total intensity. Linear segments within images show direction of polarization, the black vertical line indicates the position of the core, A0, and red circles mark positions of knot K20; 1~mas corresponds to 7.29~pc.  %7.12~pc. 
  Note that the polarized flux intensities  on May 10 and June 6 are multiplied by factors of 3 and 5, respectively, to reveal locations of peaks of polarized flux intensity, which
  are 15 mJy/beam and 9.5 mJy/beam, respectively, for the May 10 (MJD=58979) and June 6 (MJD=59006) images.}
\label{fig:vlba7mm}
\end{figure}

The total and polarized intensity VLBA maps of \src\ are presented in Fig.~\ref{fig:vlba7mm}. 
The images exhibit the bright VLBI core located at the north-western end of the jet and a weak extended jet at position angle $\sim130^\circ$.

The total and polarized intensity images are modeled by circular components with Gaussian brightness distributions. 
Two bright features are apparent at each epoch: the core A0, and a knot K20. We assume that the core is a stationary feature
of the jet and calculate parameters of K20 with respect to the core.  Parameters of the modeling are given in Table~\ref{tab:vlbamod}. 
\begin{table*}[]
    \centering
    \begin{tabular}{l|c|r|r|r|r|r|r}
    \hline
MJD & Knot &$S$ & $R$ & $\Theta$ & $a$ & P & EVPA \\
(1) & (2) & (3) & (4) & (5) & (6) & (7) & (8) \\
\hline 
58916&A0&0.90$\pm$0.06&0.0&\nodata&0.029$\pm$0.007&3.5$\pm$0.3&110$\pm$6\\
58979&A0&0.52$\pm$0.06&0.0&\nodata&0.023$\pm$0.007&1.5$\pm$0.5&$-$20$\pm$8\\
59006&A0&0.44$\pm$0.05&0.0&\nodata&0.066$\pm$0.010&0.8$\pm$0.3&78$\pm$11\\\hline
58916&K20&0.46$\pm$0.04&0.073$\pm$0.018&118$\pm$9&0.088$\pm$0.014&6.5$\pm$0.4&113$\pm$7\\
58979&K20&0.26$\pm$0.03&0.118$\pm$0.024&120$\pm$8&0.096$\pm$0.017&5.7$\pm$0.3&114$\pm$6\\
59006&K20&0.10$\pm$0.02&0.151$\pm$0.015&117$\pm$8&0.067$\pm$0.015&8.3$\pm$0.5&91$\pm$6\\
\hline
\end{tabular}

\caption{
Parameters of core A0 and knot K20. 
Columns of table are: (1) epoch of observation in MJD; (2) designation of knot; (3)  flux density, $S$, in Jy; (4) distance from the core, $R$, in mas; (5) position angle with respect to the core, $\Theta$, in degrees; (6) FWHM size of knot, $a$, in mas; (7) degree of polarization, $P$, in \%; and (8) position angle of polarization, EVPA, in degrees. For model parameters, 1$\sigma$ uncertainties are provided.
}
\label{tab:vlbamod}
\end{table*}

\begin{figure}%[tp]
  \centering
\includegraphics[width=0.49\textwidth]{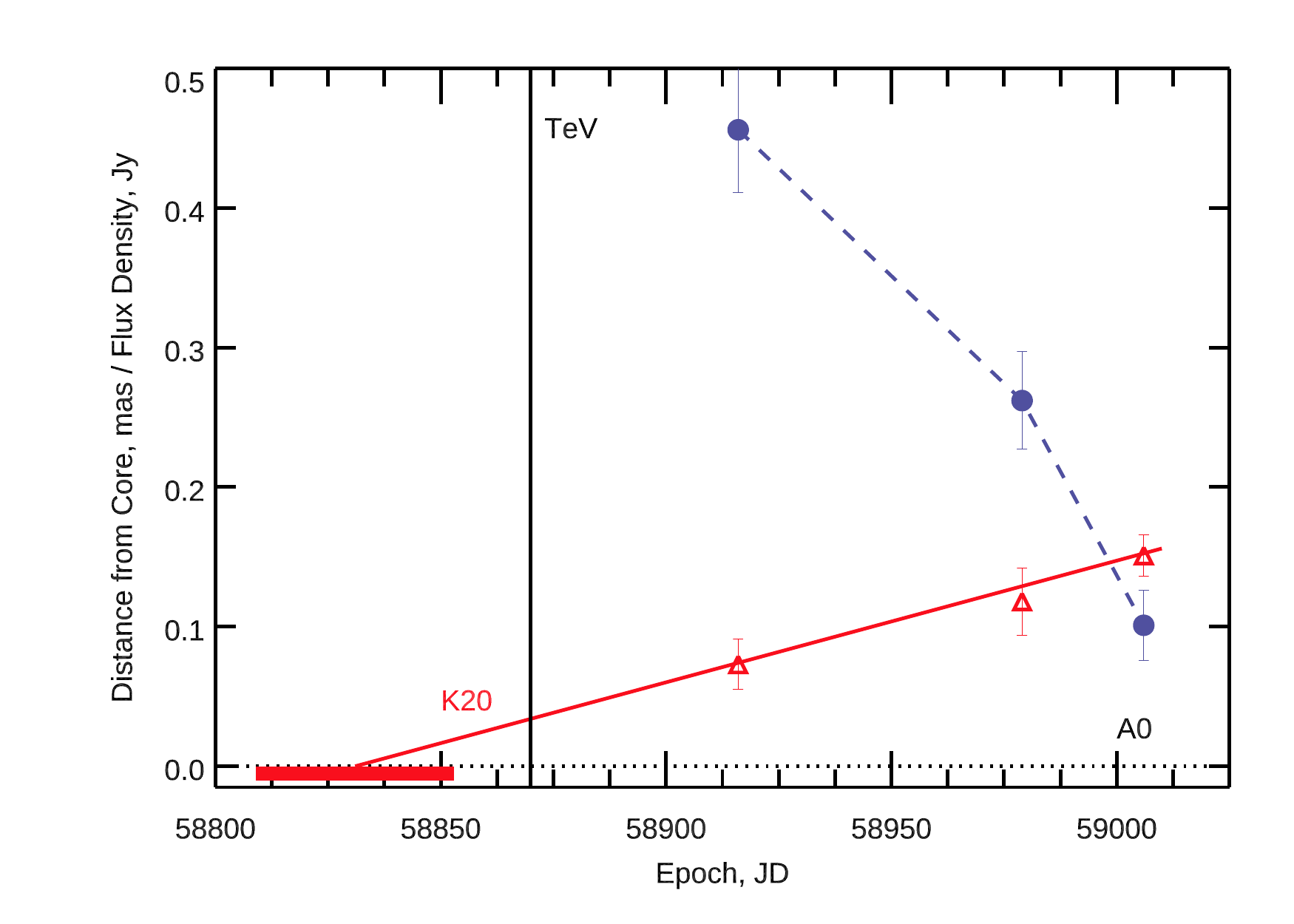}  
  \caption{Separation of K20 from the core with time (red triangles) according to
  modeling; the red line represents a linear approximation of the motion. Blue circles and dashed line correspond to the light curve of K20 at 43~GHz; the black vertical line indicates the time of the VHE event. 
 \label{fig:vlbaK20}
}
\end{figure}

According to Fig.~\ref{fig:vlba7mm} and Table~\ref{tab:vlbamod}, knot K20 is the most
polarized feature of the jet. Fig.~\ref{fig:vlbaK20} shows the separation of K20
from the core as a function of time. According to a linear approximation, K20 moves with a proper motion $\mu$ = 0.30$\pm$0.04~mas/yr, which translates into superluminal motion down the jet with apparent speed $\beta_{\text{app}}=12.0\pm1.7$  %$\beta_{\text{app}}=11.9\pm1.7$ 
in units of $c$.
Such motion suggests that the ejection\footnote{passage of the centroid of K20 through the centroid of A0} of K20 through the VLBI core occurred on MJD 58831$\pm$21 (2019 December 13). 
Fig.~\ref{fig:vlbaK20} also shows the light curve of K20, which reveals a very fast
decrease of the knot flux density by a factor of $\sim$4 in three months, which 
corresponds to the timescale of variability of 0.16$\pm$0.03~yr, as defined according to \cite{BJO74}. Applying the formalism proposed by \cite{J05}, this timescale of variability and the average size of K20 (0.084$\pm$0.015\,mas according to Table~\ref{tab:vlbamod}) give a value of the Doppler factor of K20 $\delta$=33$\pm$9.
The latter, along with the apparent speed of the knot, allow us to estimate the
Lorentz factor, $\Gamma_{\text b}$=19$\pm$9, and viewing angle $\Theta_\circ$=1.1$^\circ\pm$0.6$^\circ$ of K20. 
Using the proper motion and size of K20, we can estimate that during the time of VHE emission the upstream edge of the knot was passing through the centroid of the core. The direction of polarization in K20 is close to the jet direction, which
suggests that K20 is most likely a moving shock whose surface is oriented transverse to the jet axis.  This is supported by a higher degree of polarization of the knot compared with the core, which implies ordering of the magnetic field as expected by  such a shock. 

\section{Spectral energy distribution modeling} \label{sec:model}
FSRQ gamma-ray emission is usually explained in the framework of an external Compton (EC) model. 
Moreover, the optical spectroscopy measurements performed during the dimming phase of the emission (see Fig.~\ref{fig:FeIIfit}) show a significant increase in flux of FeII + FeIII lines. 
Since the source emits up to VHE range, the most natural target for EC process is the dust torus (DT) radiation field (see, e.g., \citealp{co18, be19}). 
DT radiation field, contrary to BLR would not absorb strongly the sub-TeV gamma-rays, and thus is the common assumption for the dominating radiation field in the modeling of FSRQs detected at VHE gamma rays. 
On the other hand the large increase of the optical flux during the gamma-ray flare can provide as well a significant target for  synchrotron-self-Compton (SSC) process.
We therefore investigate a scenario in which both SSC and EC processes are possible. 

Intriguingly, compared to the low-state spectrum, the investigated synchrotron and gamma-ray peaks are shifted towards higher energies. 
Similar behavior has been observed also in a few other FSRQs during enhanced states, e.g.: PMN J2345--1555  \citep{gh13}, 4C $+$49.22 \citep{cu14}, PKS 1441$+$25 \citep{ah15,ab15}, PKS 1510--089 \citep{dammando11}, and PKS 0346--27 \citep{ang19b}.
For most of these cases, however, the peak frequency did not reach beyond $10^{14}$\,Hz.
Therefore, the behavior observed in \src{}, in particular the SED peaking at a few times $10^{14}$\,Hz in optical range during period B, while not being unique is still rarely observed in FSRQs. 
The peak position traces the electron energy distribution (EED), however, it is also dependent on other physical parameters of the source (e.g. on the beaming). 

We model the source in a framework of a simple one-zone model in which a spherical emission region is homogeneously and isotropically filled with an electron distribution and magnetic field. 
%We consider a broken power-law energy distribution of electrons with a slope $p_1$ between $\gamma_{\rm min}$ and $\gamma_{\rm b}$, and $p_2$ above $\gamma_{\rm b}$ until $\gamma_{\rm max}$. 
We consider a broken power-law energy distribution of electrons, i.e. 
$dN_e/d\gamma\propto \gamma^{-p1}$ for $\gamma_{\rm min}<\gamma<\gamma_{\rm b}$ and 
$dN_e/d\gamma\propto \gamma^{-p2}$ for $\gamma_{\rm b}<\gamma<\gamma_{\rm max}$. 
The electrons in the blob are also exposed to an additional, directional radiation field coming from the DT. 
The model assumes ring geometry of the DT, and thus depends on the distance of the emission region from the black hole. 
The SED model was generated with \texttt{agnpy}\footnote{\burl{https://github.com/cosimoNigro/agnpy/}} \citep{agnpy}, which implements the synchrotron and Compton processes following the prescriptions described in \citet{dm09,fi16}. 
We fix the Lorentz and Doppler factors of the blob to $\Gamma=40$ and $\delta=40$, respectively. 
Note that those values are somewhat larger (in particular the $\Gamma$ factor of the blob) than the jet parameters estimated using VLBI observations (see Section~\ref{sec:vlbi}). 
However, the VLBI measurements are performed a few months after the flaring event and trace the later evolution of the blob, thus some change of parameters of the jet might have happened in the meantime (in particular deceleration). 
The size of the blob is limited by the variability condition. 
The values that we use in the modeling, $r_b=3 - 6 \times 10^{16}$\,cm, correspond to the light crossing time of 12-24\,hrs, of the order of the time scale of the observed variability.
We assume that the emission region is located at the distance of $d=2.5 \times 10^{18}$\,cm, i.e. of the order of $\sim \Gamma r_b$. 
We use the accretion disk luminosity $L_{\rm disk}=2 \times 10^{46} \mathrm{erg\,s^{-1}}$ (see Section~\ref{sec:spectro}) to estimate the size of the BLR and DT following the scaling relations of \cite{gt09}.
Note that while the optical spectra used in this estimation are not fully simultaneous with the broadband emission data used for the modeling, the size of the DT makes the emission quasi-stable at the time scales of years. 
DT is simulated as a thin ring with a radius of $R_{\rm DT}=1.1\times 10^{19}$\,cm ($3.6$\,pc).  % obtained from : $R_{\rm DT}=2.5\times 10^{18} (L_{\rm disk}/10^{45} \mathrm{erg\,s^{-1}})^{1/2} \mathrm{cm}$.
Since the estimated size of the BLR is $R_{\rm BLR}=4.4\times 10^{17}$\,cm ($0.14$\,pc), the emission region is not affected by the BLR but it is deep in the DT radiation field. 
We assume that 0.6 of the disk luminosity is reprocessed in the DT radiation.

In order to obtain EED in a self-consistent way we introduce the acceleration parameter $\xi$, defined such that acceleration gain of electrons is 
$(dE/dt)_{\mathrm{acc}} = \xi c E / R_L$, where $c$ is the speed of light and $R_L$ is the Larmor radius. 
The maximum energy of the electrons is obtained from comparing the acceleration energy gain with energy losses due to  IC (in Thomson regime) cooling:
\begin{equation}
    \gamma_\mathrm{max}=\sqrt{3 \xi  e  B / 4 \sigma_T  u_{ph}'}, \label{eq:gmax}
\end{equation}
where $B$ is the co-moving magnetic field in the blob, $e$ is elementary charge, $\sigma_T$ is the Thomson cross-section. 
$u_{ph}'$ is the co-moving energy density of the dominating radiation field. 
In the case of the parameter sets used in the modeling the dominating radiation field is originating from the DT, however we explicitly check also possible limit from the SSC process.  
The maximum electron energies are also tested against the dynamical time scale, by comparing the acceleration time scale with the ballistic time scale\footnote{also called light crossing time scale}, $T_\mathrm{bal}\simeq r_b/c$ of the blob crossing its size, and against synchrotron energy losses. However, neither of the two processes is dominant in the cases investigated here.
Since during $T_\mathrm{bal}$ the blob crosses its size, it might escape the region in the jet (e.g. a stationary shock) where efficient acceleration occurs. 
Note that $T_\mathrm{bal}$ determines also the time scale of adiabatic losses of the electron (at the assumption that the blob distance is $d=r_b\Gamma$). 

\begin{figure*}%[t!]
   \centering
  \includegraphics[width=0.45\textwidth, trim=5 10 50 40,clip]{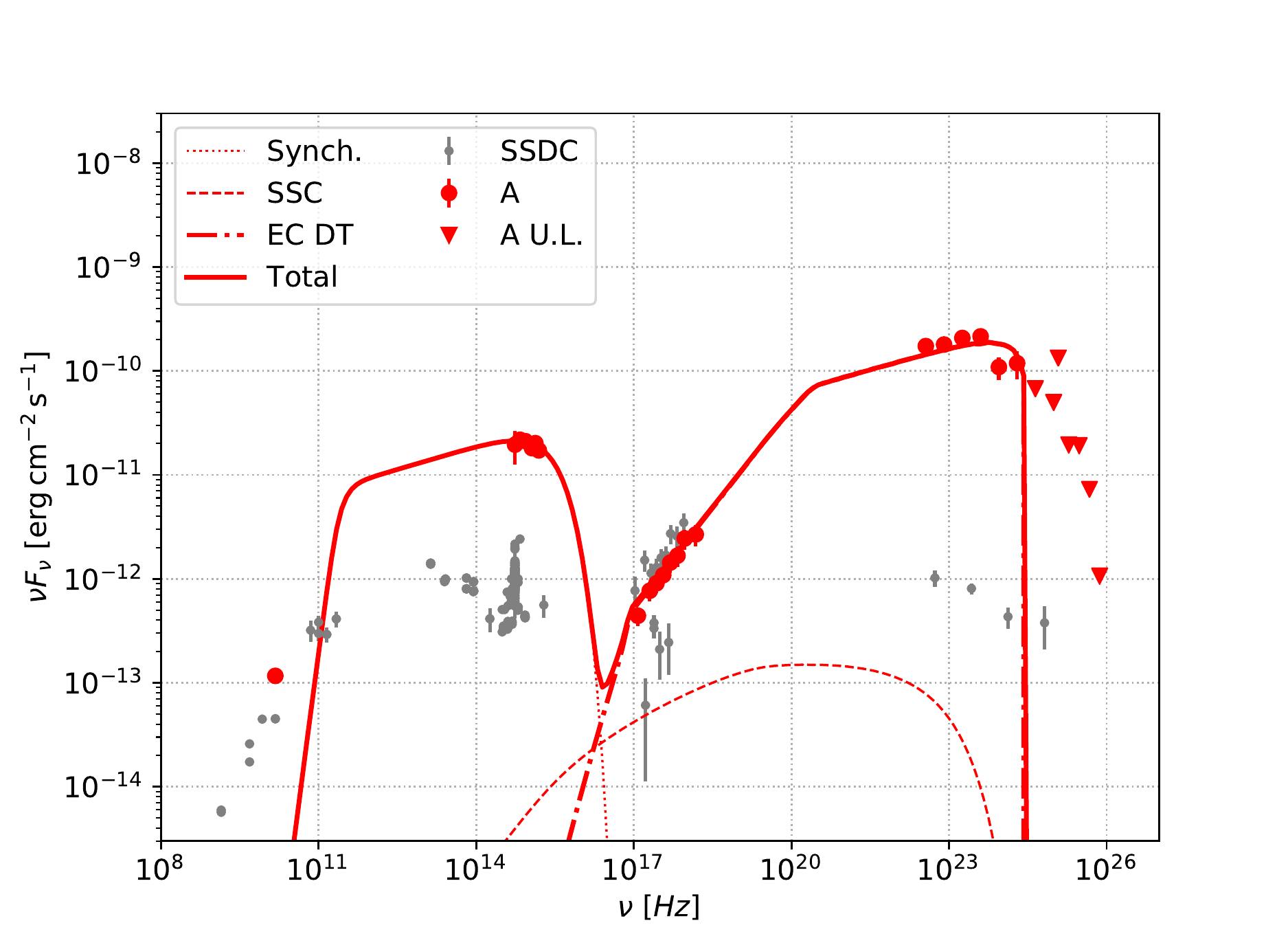}
  \includegraphics[width=0.45\textwidth, trim=5 10 50 40,clip]{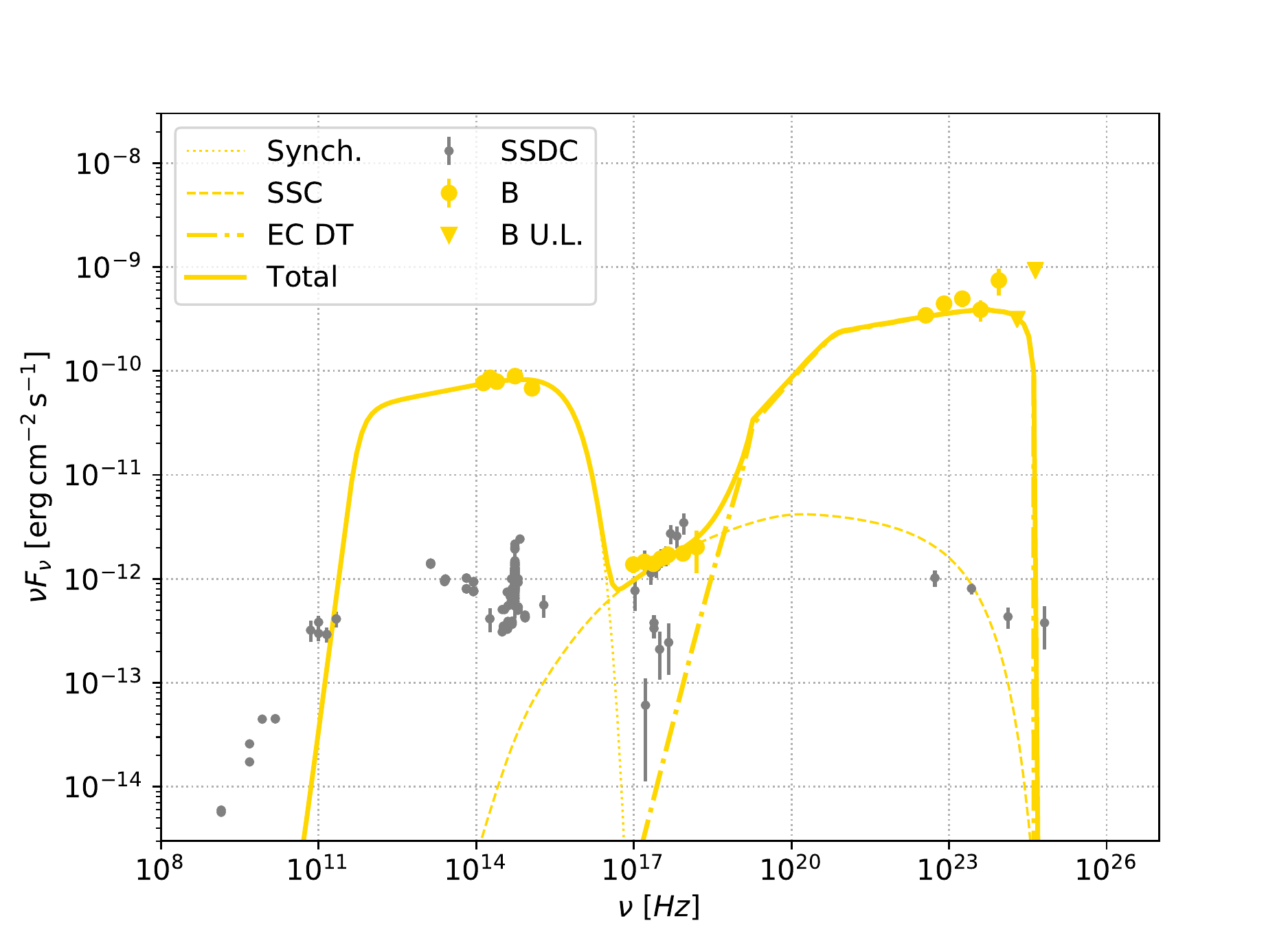}\\
  \includegraphics[width=0.45\textwidth, trim=5 10 50 40,clip]{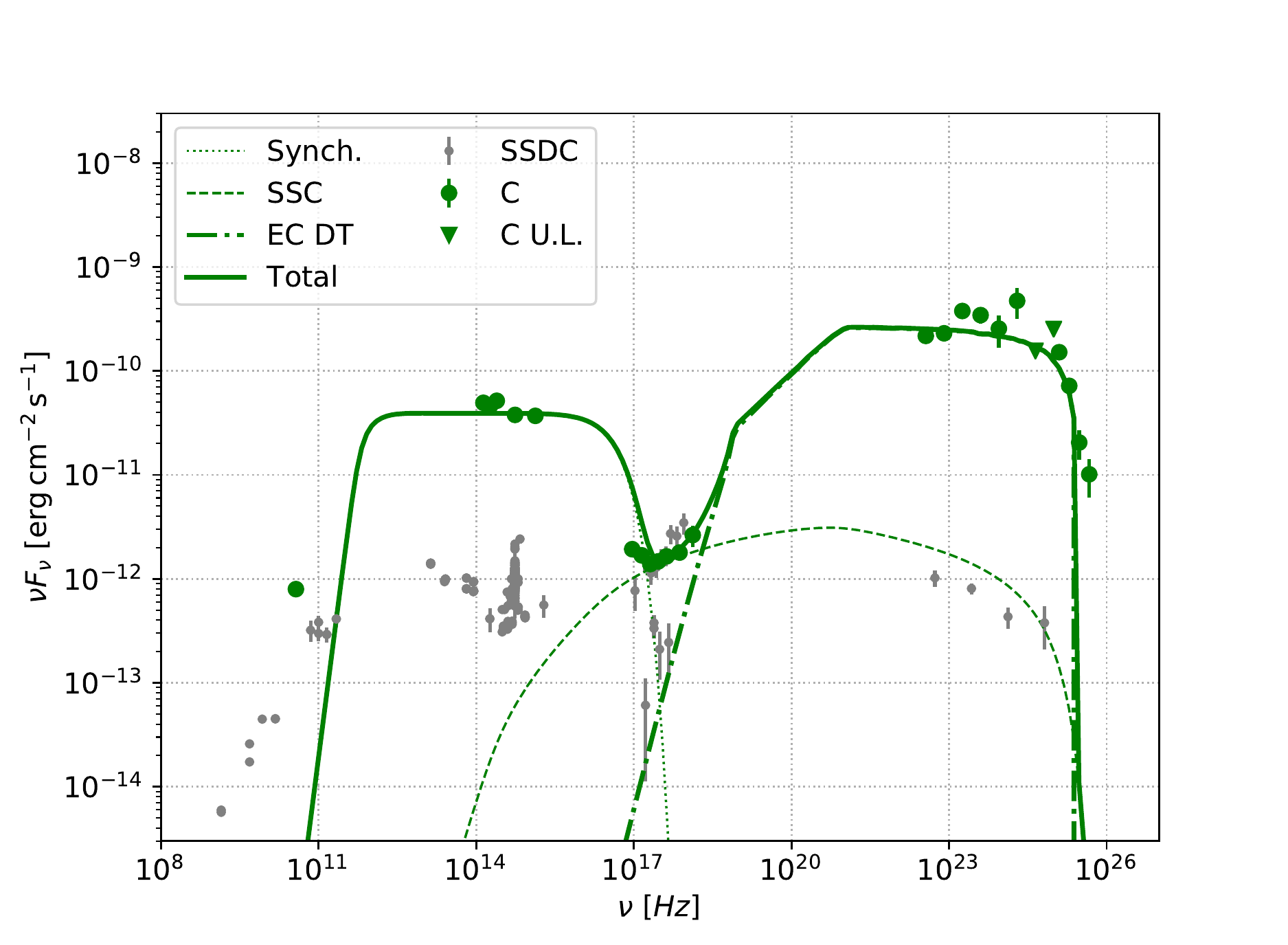}
  \includegraphics[width=0.45\textwidth, trim=5 10 50 40,clip]{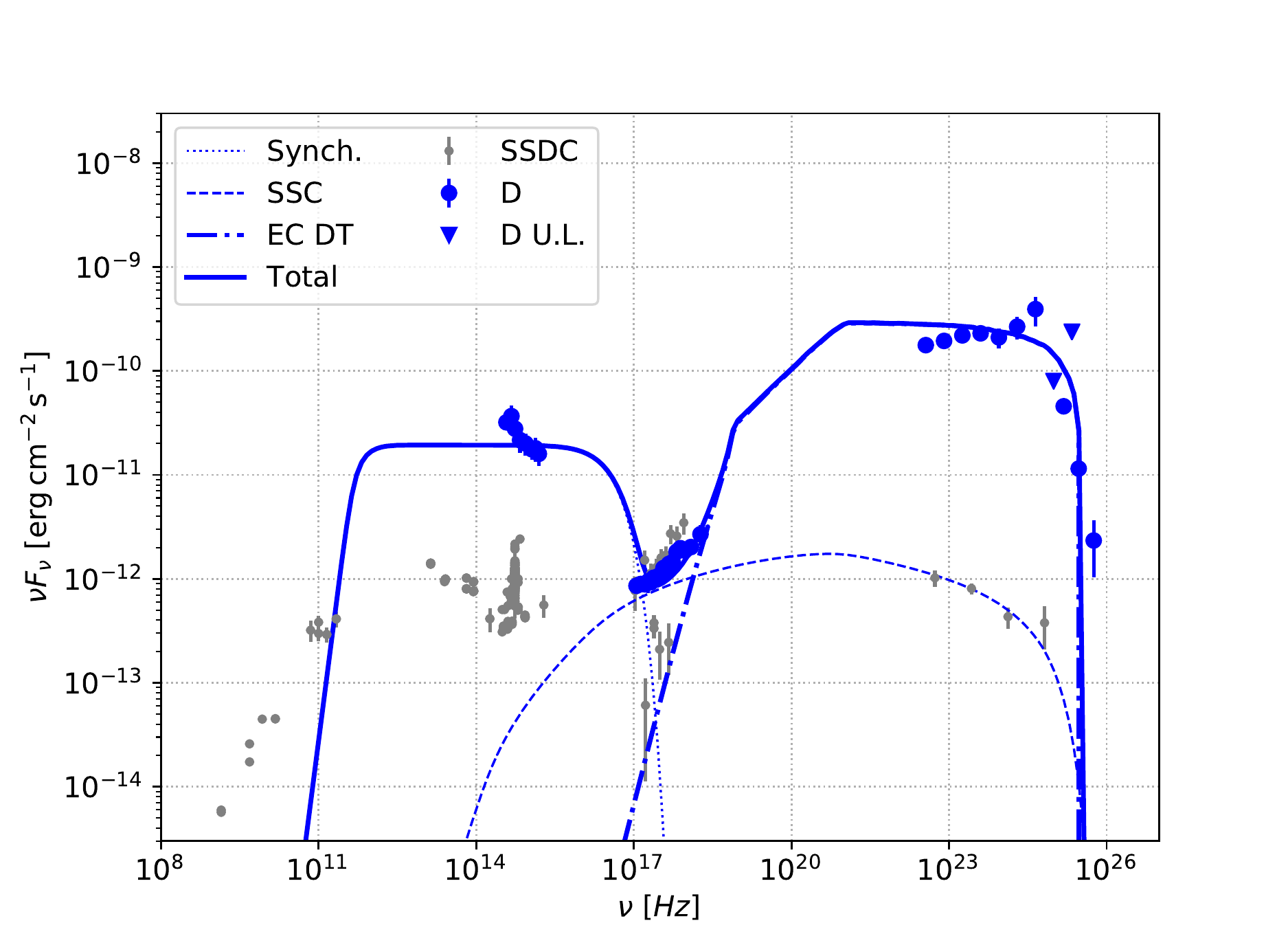}
  \caption{ Multiwavelength SED of \src\ in the four periods A -- before the flare, B -- optical flare, C -- VHE gamma-ray flare, D -- after the flare and archival data (gray). 
    Different radiation processes are shown with different line styles: dotted lines -- synchrotron, dashed -- SSC, dot-dashed -- EC, solid -- sum of components.  
    Model lines are corrected for EBL absorption according to \citet{do11}. }  
  \label{fig:mwl_sed_model}
\end{figure*}

\begin{figure*}%[t!]
   \centering
  \includegraphics[width=0.45\textwidth]{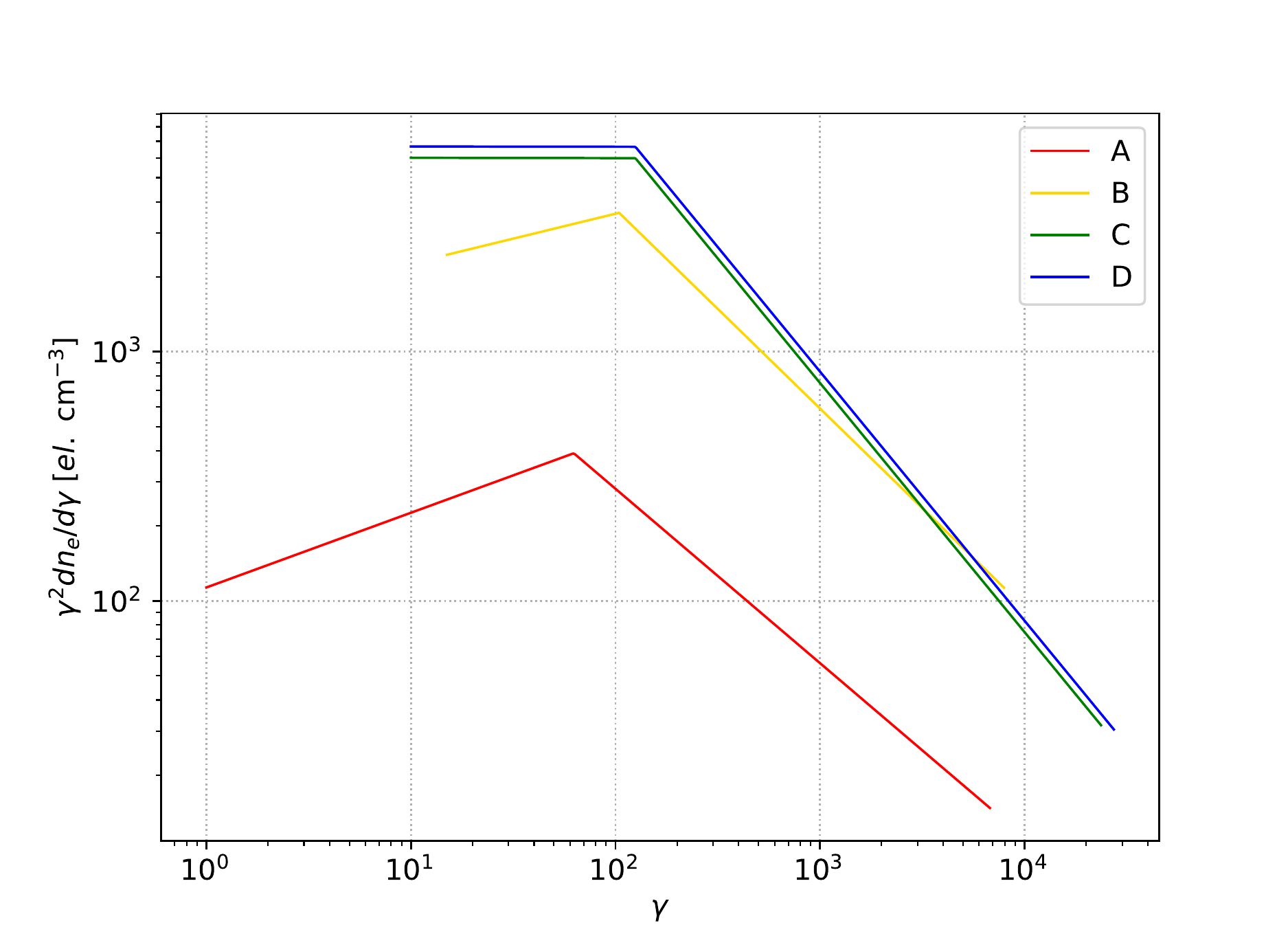}
  \includegraphics[width=0.45\textwidth]{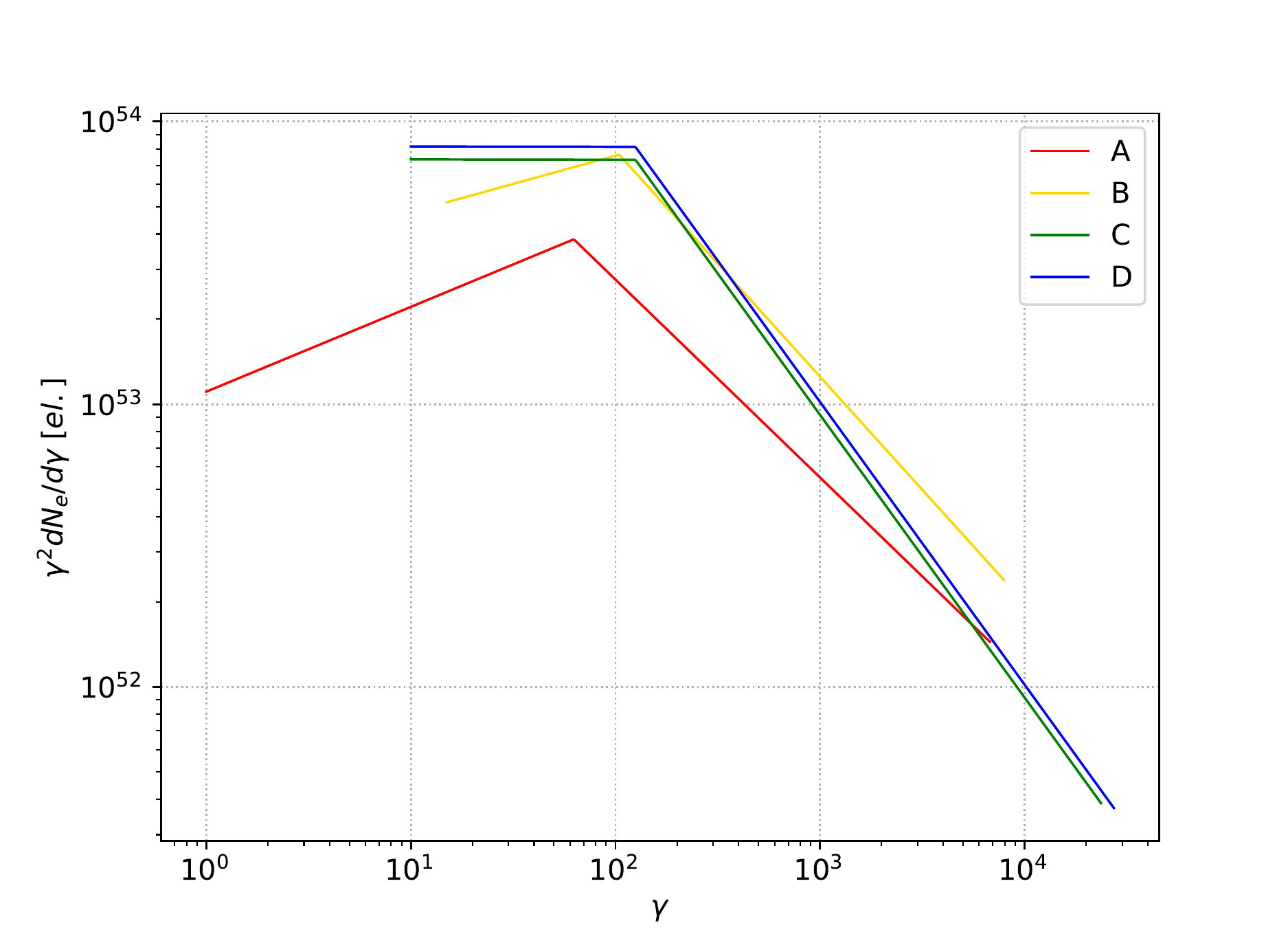}
  \caption{ 
  Evolution of the EED in the frame of the blob: energy density (left) and total energy integrated over the blob (right).  Different line colors represent different stages of the flare.}  
  \label{fig:eed_evol}
\end{figure*}

We also introduce a canonical cooling break (steepening of EED by 1) at the energy where the dominating cooling time scale is equal to the ballistic time scale, $T_\mathrm{bal}$:
\begin{equation}
    \gamma_\mathrm{b} = 3 m_e c^2  / 4 \sigma_T  u_{ph}' r_b, \label{eq:gbreak}
\end{equation}
where $m_e$ is the electron mass.
In order not to overshoot the X-ray flux with IC photons and still be able to explain the gamma-ray emission during the periods B, C and D it is necessary to assume that the EED starts at $\gamma_{\rm min}>1 $. 
We apply values of $\gamma_{\rm min}=10-15$ in the modeling, while for period A we use $\gamma_{\rm min}=1$. 
We then tune the magnetic field $B$ and the energy density of the electrons $u'_e$ to reproduce the levels of synchrotron and IC emission.
The index of EED before the break is selected to roughly reproduce the spectral shape in IR-UV (except of period D) and gamma-ray bands. 
The resulting value $p_1=1.7-2$, is close to the canonical non-relativistic value of $2$ (see, e.g., \citealp{dr83}). 
The location of the valley between the peaks is most sensitive to the maximum $\gamma$ factor of the electrons and the onset of the IC peak. 
The maximum energies of the electrons are tuned by the acceleration parameter $\xi$ (see Eq.~\ref{eq:gmax}) with an additional constraint from the VHE gamma-ray spectrum. 
The depth of the valley is modified additionally by the SSC component. 
By tuning the compactness of the blob (i.e. varying the radius of the blob within a factor of 2 with simultaneous keeping the same total power in electrons that fixes the synchrotron and EC emission at roughly the same level), which affects the SSC emission,  we fit the depth of the dip in periods B, C and D. 
Since in all the studied energy bands the emission during the whole investigated period was significantly larger then the historical measurements we neglect the possible low-state emission of the source in the modeling. 
The broadband spectra obtained from the modeling are compared with measured ones in Fig.~\ref{fig:mwl_sed_model}. 
The corresponding evolution of EED is shown in Fig.~\ref{fig:eed_evol}. 
Parameters of the modeling are shown in Table~\ref{tab:table_model}. 

\begin{table*}[]
    \centering
    \begin{tabular}{c|c|c|c|c|c|c|c||c|c|c|c|c}
Period &  $\delta$ & $r_b$ [cm] & $\xi$ & $B$ [G] & $U'_e\mathrm{[10^{48} erg]}$ & $p_1$ & $\gamma_\mathrm{min}$ & $p_2$ &  $\gamma_\mathrm{break}$ & $\gamma_\mathrm{max}$  & $u'_e\mathrm{[erg\,cm^{-3}]}$ & $u'_e/u'_B$ \\ \hline
% before UVOT-recalibration
%A & 40 & $6.16 \times 10^{16}$ & $0.3 \times 10^{-7}$ & 0.70 & 1.18 & 1.7 & 1 & 2.7 & 63 & 6900 & 1.2 $\times 10^{-3}$ & 0.06\\ \hline
%B & 40 & $3.70 \times 10^{16}$ & $0.3 \times 10^{-7}$ & 0.95 & 1.76 & 1.8 & 15.0 & 2.8 & 104 & 8000 & 8.3 $\times 10^{-3}$ & 0.23 \\ \hline
%C & 40 & $2.03 \times 10^{16}$ & $3.0 \times 10^{-7}$ & 0.75 & 0.88 & 1.7 & 10.0 & 2.7 & 190 & 22600 & 25.0 $\times 10^{-3}$ & 1.1 \\ \hline
%D & 40 & $3.70 \times 10^{16}$ & $6.0 \times 10^{-7}$ & 0.50 & 1.18 & 1.7 & 1 & 2.7 & 104 & 26000 & 5.6 $\times 10^{-3}$ & 0.56 \\ \hline
%\\ \hline\hline
% after UVOT-recalibration
A & 40 & $6.16 \times 10^{16}$ & $0.3 \times 10^{-7}$ & 0.70 & 1.18 & 1.7 & 1 & 2.7 & 63 & 6900 & 1.2 $\times 10^{-3}$ & 0.06 \\ \hline
B & 40 & $3.70 \times 10^{16}$ & $0.3 \times 10^{-7}$ & 0.95 & 1.76 & 1.8 & 15 & 2.8 & 104 & 8000 & 8.3 $\times 10^{-3}$ & 0.23 \\ \hline
C & 40 & $3.08 \times 10^{16}$ & $3.0 \times 10^{-7}$ & 0.83 & 2.12 & 2.0 & 10 & 3.0 & 125 & 23700 & 17.3 $\times 10^{-3}$ & 0.63 \\ \hline
D & 40 & $3.08 \times 10^{16}$ & $6.0 \times 10^{-7}$ & 0.55 & 2.35 & 2.0 & 10 & 3.0 & 125 & 27300 & 19.2 $\times 10^{-3}$ & 1.6 \\ \hline

    \end{tabular}
    \caption{Parameters used for the modeling: Doppler factor $\delta$ ($\Gamma=\delta$ is assumed), co-moving size of the emission region $r_b$, acceleration parameter $\xi$, magnetic field $B$, total energy of electron $U'_e$, EED: slope before the break: $p_1$, minimum Lorentz factor $\gamma_\mathrm{min}$, slope after the break $p_2$,  the Lorentz factor of the break $\gamma_\mathrm{break}$, maximum Lorentz factor $\gamma_\mathrm{max}$, electron energy density $u'_e$, energy equipartition $u'_e/u'_B$.
    Free parameters of the model and derived parameters are put on the left and right side of the double vertical line respectively}
    \label{tab:table_model}
\end{table*}

It should be stressed that, mainly due to fixing the Doppler factor, those parameters are not uniquely determined (see, e.g., \citealp{ah17}). 
For example, assuming $\delta$ and $\Gamma$ motivated by VLBI measurements, a similar fit to the data can be achieved, however, with the size of the emission region compressed by about an order of magnitude in periods A, C and D. 
They allow us however to trace the relative evolution of the parameters with the assumption that the beaming did not change during the flare. 
The gamma-ray spectrum in the modeling is reproduced by nearly solely the EC process. 
The high spectral variability of the X-ray emission is naturally explained by three processes that contribute to it: mainly the EC and SSC emission, with a partial component from the highest-energies synchrotron radiation during period C. 

Due to its simplicity the modeling has some caveats. 
The radio emission is underestimated due to pronounced synchrotron self absorption in the model curve. 
Such emission is often attributed to a larger-scale jet, rather then the blob. 
It should be noted however that during the 2020 high state the radio emission was at higher level than in historical measurements, and also the emission has shown some variability, therefore it should be also at least partially associated with the high state. 
It is plausible that, due to evolution of the flare, low-energy electrons from the blob escape to the large scale jet without being cooled completely. 
Such an enhancement of the EED in the large scale jet during the high state would explain higher radio emission. 
In addition such escape of high energy particles along the jet would naturally explain the new component appearing in VLBI follow-up of the flare. 

In period C (and partially also in period D) the X-ray data show a clear valley between the two peaks. 
The shape of the low-energy (i.e. falling) part of the valley in this period is not fully reproduced by the model. 
This part of the SED is strongly dependent on the shape of the high-energy tail of the EED which is also constrained by the highest energy gamma rays. 
The simplifications in the modeling (homogeneous region, no non-stationary processes modifying EED within the considered period, resulting sharp cut-off of the EED) do not allow us to realistically model the full shape of the valley. 
The SED in this period might be also affected by fast variability.

%The shape of the optical spectrum is not accurately modeled in all the cases. 
%In particular in period C the NIR data are slightly overshooting the model suggesting a softer spectrum. 
%However in this period fast decay of the optical and IR emission is seen (see Fig.~\ref{fig:mwl_lc}) and the NIR data are taken before the optical ones and hence likely represent somewhat higher emission of the source.
%Similarly in period D the optical emission overshoots by a factor of two the extrapolation of UV data to lower energies. 
%This again might be connected with the variability of the source, or with additional emission of the population of partially cooled particles from the previous phases of the flare.

The slope of the spectrum is not accurately modeled in all the cases. 
In particular in period D (and partially also in period C) the optical range would require softer electron distribution than gamma-ray range. 
In particular in period C the NIR data are slightly overshooting the model suggesting a softer spectrum. 
%However in this period fast decay of the optical and IR emission is seen (see Fig.~\ref{fig:mwl_lc}) and the NIR data are taken before the optical ones and hence likely represent somewhat higher emission of the source.
%Similarly in period D the optical emission overshoots by a factor of two the extrapolation of UV data to lower energies. 
This might be connected with fast variability of the source, or with additional emission of the population of partially cooled particles from the previous phases of the flare.

Despite those caveats it is interesting to see that the obtained model parameters provide a self-consistent description of the main features of the emission during different phases of the flare. 
Comparing period B to period A modeling suggests a compression of the emission region coincident with the increase of the minimum Lorentz factor of the electrons and a mild increase of the magnetic field density and the total energy stored in electrons. 

The increased $\gamma_{\rm min}$ factor needed in the modeling (see also \citealp{ka06}) might point to a two-stage acceleration process. 
First, injection of particles with such minimum Lorentz factor has to occur, e.g., due to acceleration in a small potential drop due to reconnection of magnetic fields (see, e.g., \citealp{la15,co19}). 
Then a second process (e.g., Fermi second order acceleration) would boost the particles to a power-law spectrum up to maximal gamma factors of $\gamma_{\rm max}$. 
The acceleration coefficient $\xi$ has a rather small value, of the order of  $10^{-7}$. 
The value of $\xi$ in the modeling increases by an order of magnitude in period C %(coincident with further slight compression of the emission region and drop in the total energy stored in electrons) 
in order to explain the VHE gamma-ray emission. 
Such small values are needed to saturate the acceleration process with EC energy losses in order not to increase too much $\gamma_{\rm max}$ and in turn not to overshoot the X-ray emission by synchrotron component. 
As shown in the modeling such acceleration would be still efficient enough to explain the observed optical and HE flare. 
A natural explanation for such low values of $\xi$ is a second order Fermi process with non-relativistic scattering centers accelerating electrons in the emission region.
During the period D the VHE emission requires a further small increase of the $\xi$ parameter coincident with lower magnetic field in order not to overshoot the soft X-ray flux with the synchrotron component.

\section{Discussion and conclusions} \label{sec:conc}

Observations of \src\ with MAGIC during the enhanced state allowed us to add a new member to the sparse family of the FSRQs emitting in VHE range. 
The observations were performed during an impressive flare, with the flux of both SED peaks enhanced by about two orders of magnitude with respect to the low state. 
Monitoring observations of the source and a massive MWL campaign provided us a dataset that was used to trace the evolution of the EED during the flare. 
Interestingly, the synchrotron spectrum in the optical range during the flare (in particular period B) is hard. 
Comparing to historical data, both low- and high-energy peaks shifted by at least two orders of magnitude in frequency, such large shift being rare for a FSRQ object. 
Shifts of the peaks towards higher energies during high states is a behavior commonly observed in a sister class of objects -- BL Lacs. 
The spectra are Compton dominated, which is typical for FSRQs. However, the dominance during the peak of the flare is just a factor of a few. 

Similarly to other VHE-detected FSRQs, we get a satisfactory description of the two broad peaks of the SED as the synchrotron and EC on DT radiation field. 
The valley between the peaks is well constrained by the X-ray data, and we use it to track the changes of the compactness of the blob during the evolution of the high state. 
The modeling scenario is self-consistent in the sense of the shape of the EED being determined by the balance of acceleration and cooling processes. 
The variability of the emission between different phases of the enhanced state is explained mainly by a combination of variations of the compactness of the emission region, the minimal injection energy of electrons and increase of the acceleration parameters. 
In addition to achieve a satisfactory fit coincident small variations of the magnetic field, total energy stored in electrons, and injection slope has been assumed. 

%The optical spectroscopy observations show a prominent MgII line, which we used to estimate the accretion disk luminosity at $2\times 10^{46} \,\mathrm{erg\, s^{-1}}$.
The optical spectroscopy observations revealed a prominent MgII line, that does not show flux variability exceeding the uncertainties of the measurements. 
We explain this as the line being produced within a canonical BLR, and so has a much longer timescale of variability.
Therefore, the MgII line is a good proxy for estimating the accretion disk luminosity at $2\times 10^{46} \,\mathrm{erg\, s^{-1}}$.   
Additionally, a broad FeII bump has been observed, with the luminosity increasing with the increase of the optical continuum emission. 
The fast variability of the FeII bump suggests that it originates in a much smaller region (possibly located close to the jet axis) than the regular BLR. 
Moreover, since the flux of FeII bump correlates with the synchrotron continuum, the bump should be produced farther from the black hole in the vicinity of the jet. 
Additionally, the shift of the bump to the blue side could be explained if it is produced in a wind surrounding the jet. 
This suggests a possible interaction of the jet with a FeII-emitting cloud.

Optical polarimetry, that started a few days after optical peak, shows a very low polarization and smooth EVPA rotation. 
This makes \src\ another FSRQ in which VHE gamma-ray emission is detected contemporaneous to EVPA rotations. 
Intriguingly, the VLBI observations in the follow up of the flare show an emission of a superluminal radio knot contemporaneous with the high gamma-ray state. 
Similar association for VHE gamma-ray emission + EVPA rotation + VLBI component ejection has been previously suggested also for another FSRQ PKS\,1510$-$089 \citep{ah17a}. 

The detection of \src{} in the VHE gamma-ray range and the extensive monitoring campaign during this event add another piece to solving the puzzle of the origin of the highest energy emission of FSRQ objects. 
It is interesting that some of the emission features associated with these observations of \src{} are similar to observations of other FSRQs, in particular of PKS\,1510$-$089, so far the most thoroughly studied FSRQ in the VHE range. 
\src{} might be a cousin of PKS 1510$-$089, twice as distant but intrinsically more luminous.

\begin{acknowledgements}
\\
\emph{We would like to dedicate this paper to the memory of our friend and
colleague, Dr. Valeri Larionov (1950-2020), who enthusiastically contributed to
this and many other projects aimed at understanding blazars.}\\
We would like to thank the Instituto de Astrof\'{\i}sica de Canarias for the excellent working conditions at the Observatorio del Roque de los Muchachos in La Palma. The financial support of the German BMBF and MPG; the Italian INFN and INAF; the Swiss National Fund SNF; the ERDF under the Spanish MINECO (FPA2017-87859-P, FPA2017-85668-P, FPA2017-82729-C6-2-R, FPA2017-82729-C6-6-R, FPA2017-82729-C6-5-R, AYA2015-71042-P, AYA2016-76012-C3-1-P, ESP2017-87055-C2-2-P, FPA2017-90566-REDC); the Indian Department of Atomic Energy; the Japanese ICRR, the University of Tokyo, JSPS, and MEXT; the Bulgarian Ministry of Education and Science, National RI Roadmap Project DO1-268/16.12.2019 and the Academy of Finland grant nr. 320045 is gratefully acknowledged. This work was also supported by the Spanish Centro de Excelencia ``Severo Ochoa'' SEV-2016-0588 and SEV-2015-0548, the Unidad de Excelencia ``Mar\'{\i}a de Maeztu'' MDM-2014-0369 and the "la Caixa" Foundation (fellowship LCF/BQ/PI18/11630012), by the Croatian Science Foundation (HrZZ) Project IP-2016-06-9782 and the University of Rijeka Project 13.12.1.3.02, by the DFG Collaborative Research Centers SFB823/C4 and SFB876/C3, the Polish National Research Centre grant UMO-2016/22/M/ST9/00382 and by the Brazilian MCTIC, CNPq and FAPERJ.
% % Fermi-lAT
The \textit{Fermi}-LAT Collaboration acknowledges generous ongoing support from a
number of agencies and institutes that have supported both the
development and the operation of the LAT, as well as scientific data
analysis. These include the National Aeronautics and Space
Administration and the Department of Energy in the United States; the
Commissariat \'a l’Energie Atomique and the Centre National de la
Recherche Scientifique/Institut National de Physique Nucl\'eaire et de
Physique des Particules in France; the Agenzia Spaziale Italiana and
the Istituto Nazionale di Fisica Nucleare in Italy; the Ministry of
Education, Culture, Sports, Science and Technology (MEXT), High Energy
Accelerator Research Organization (KEK), and Japan Aerospace
Exploration Agency (JAXA) in Japan; and the K. A. Wallenberg
Foundation, the Swedish Research  Council,
and  the  Swedish  National  Space  Board in Sweden.

Additional support for science analysis during the operations
phase is gratefully acknowledged from the Istituto Nazionale di
Astrofisica in Italy and the Centre National d'Etudes Spatiales in
France. This work was
performed in part under DOE Contract DE-AC02-76SF00515.
% Metsahovi
This publication makes use of data obtained at the Mets\"ahovi Radio Observatory, operated by Aalto University in Finland. 
% OVRO
This research has made use of data from the OVRO 40-m monitoring program \cite{ri11} %(Richards, J. L. et al. 2011, ApJS, 194, 29) 
which is supported in part by NASA grants NNX08AW31G, NNX11A043G, and NNX14AQ89G and NSF grants AST-0808050 and AST-1109911.
% Perkins
This study was based in part on observations conducted using the 1.8m Perkins Telescope Observatory (PTO) in Arizona, which is owned and operated by Boston University.
% BU
The research at Boston University was supported in part
by NASA Fermi GI program grants 80NSSC17K0649, 80NSSC19K1504, and 80NSSC19K1505.  
%ASAS-SN
We thank the ASAS-SN team for making their data publicly available. 
%VLBA
The VLBA is an instrument of the National Radio Astronomy Observatory. The National Radio Astronomy Observatory is a facility of the National Science Foundation operated by Associated Universities, Inc.
%LDT
This work made use of the Lowell Discovery Telescope (formerly Discovery Channel Telescope) at Lowell Observatory. Lowell is a private, nonprofit institution dedicated to astrophysical research and public appreciation of astronomy and operates the LDT in partnership with Boston University, the University of Maryland, the University of Toledo, Northern Arizona University, and Yale University. 
% Rozhen
We acknowledge support by Bulgarian National Science Fund under grant DN18-10/2017 and National RI Roadmap Projects DO1-277/16.12.2019 and DO1-268/16.12.2019 of the Ministry of Education and Science of the Republic of Bulgaria.
\end{acknowledgements}

% 
% %%%%%%%%%%%%%%%%%%%%%%%%%%%%%%%%%%%%%%%%%%%%%%%%%%
% 
% %%%%%%%%%%%%%%%%%%%% REFERENCES %%%%%%%%%%%%%%%%%%

%%%%%%%%%%%%%%%%%%%%%%%%%%%%%%%%%%%%%%%%%%%%%%%%%%

%%%%%%%%%%%%%%%%% APPENDICES %%%%%%%%%%%%%%%%%%%%%

\appendix
\section{}
\subsection{\textit{XMM-Newton} and \swift\ short-term variations}

We have investigated short-term variability of the source in UV and X-rays by analyzing {\it Swift} and {\it XMM-Newton} data.
 
Correcting for instrumental artifacts and pile-up, no significant 
($>$ 3-$\sigma$) increase of count rate has been observed in the background-subtracted consecutive XRT observation segments.

We search also for significant  ($>$ 3-$\sigma$) changes of magnitude between two consecutive UVOT exposures collected with the same filter at the same epoch. This results in two events observed in $w1$ filter on 2020 January 19 (MJD=58867) and 31 (MJD=58879),  with a change of 0.16 mag in 27.6 ks and 0.41 mag in 34.4 ks, respectively.

\begin{figure}
  \centering
  \includegraphics[width=0.49\textwidth]{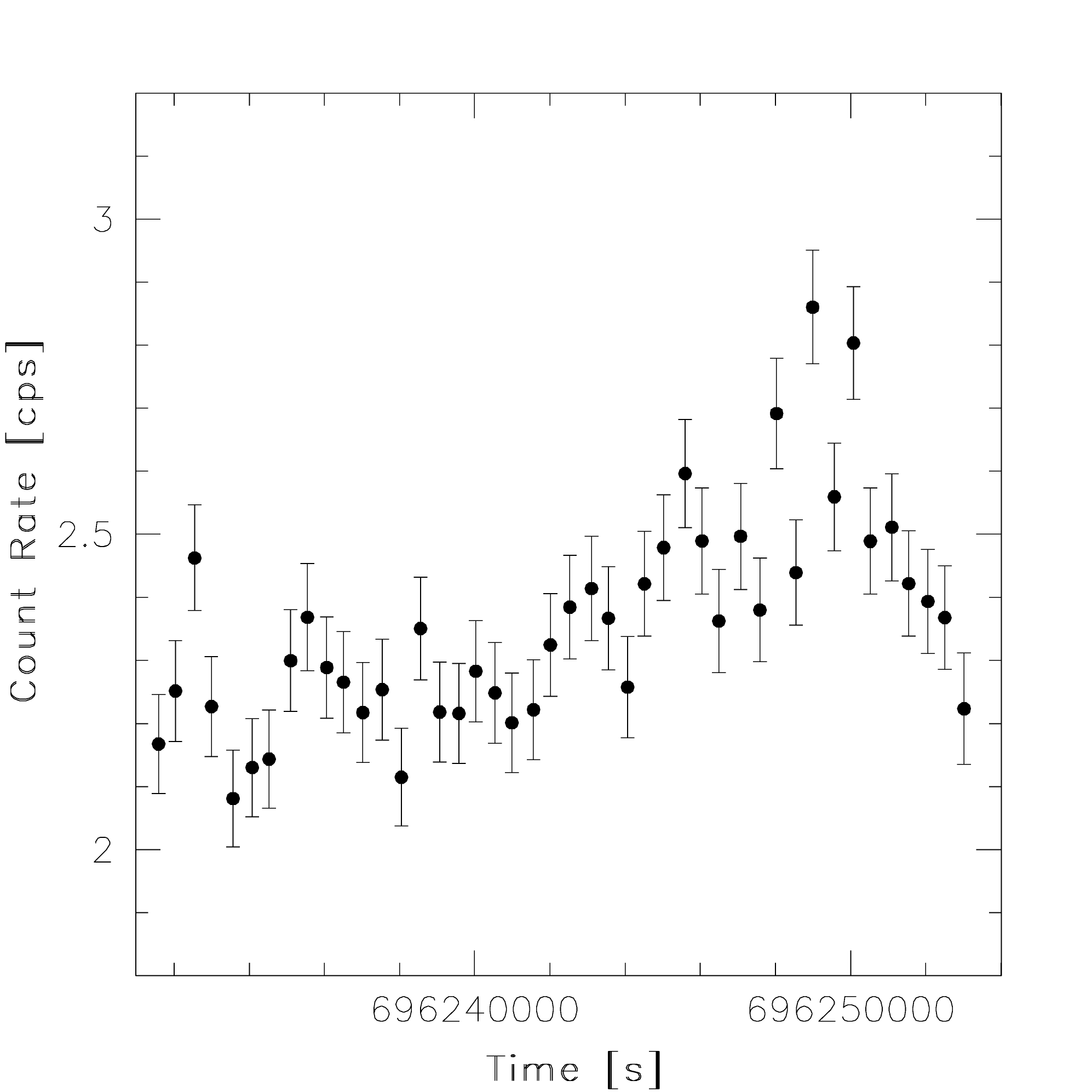}
  \caption{{\it XMM–Newton} EPIC pn light curve over 0.3--10\,keV with 500\,s bins of \src.
  \label{fig:XMM_lc}
  }  
\end{figure}

Moreover, we produce a light curve of the {\it XMM-Newton} pn count rate with bin of 500\,s (Fig. \ref{fig:XMM_lc}). The light curve shows only moderate variability, with the count rate varying between 2.08 and 2.86\,cps. The fractional variability (see \citealp{va03} for details) is 0.064 $\pm$ 0.005. A larger variability has been observed in the second part of the observation, in particular an increase of $\sim$15\% of the count rate at the time of the highest peak. 

\subsection{\swift\ spectral fits}\label{sec:swift}
In Table~\ref{XRT_1424} we report the results of analysis of all \swift\ observations, including both historical observations and the ones performed during the high state of the source shown in detail in Fig.~\ref{fig:mwl_lc}. 

\begin{table*}
\begin{center}
\begin{tabular}{ccccc}
\hline
\multicolumn{1}{c}{\textbf{Date}} &
\multicolumn{1}{c}{\textbf{MJD}} &
\multicolumn{1}{c}{\textbf{Net exposure time}} &
\multicolumn{1}{c}{\textbf{Photon index}}      &
\multicolumn{1}{c}{\textbf{Flux$_{\rm\,0.3-10\,keV}$}}  \\
\multicolumn{1}{c}{(UT)} &
\multicolumn{1}{c}{} &
\multicolumn{1}{c}{(s)} &
\multicolumn{1}{c}{($\Gamma_{\rm\,X}$)}  &
\multicolumn{1}{c}{(10$^{-12}$ erg cm$^{-2}$ s$^{-1}$)}  \\
\hline
2018-02-22      &    58171.915197  &   2872  &  1.49 $\pm$ 0.22 & 1.99 $\pm$ 0.37 \\
2018-12-12      &    58464.132329  &   1818  &  1.51 $\pm$ 0.25 & 2.76 $\pm$ 0.54 \\
2018-12-14      &    58466.253218  &   2110  &  1.37 $\pm$ 0.18 & 4.15 $\pm$ 0.65 \\
2018-12-16      &    58468.386537  &   1943  &  1.50 $\pm$ 0.21 & 3.22 $\pm$ 0.54 \\
2019-06-25      &    58659.880024  &   1975  &  1.89 $\pm$ 0.21 & 3.18 $\pm$ 0.45 \\
%2019-06-27/29   &    58662.937172  &   2008  &  1.25 $\pm$ 0.24 & 3.07 $\pm$ 0.64 \\
2019-06-27/29   &    58663  &   2008  &  1.25 $\pm$ 0.24 & 3.07 $\pm$ 0.64 \\
2019-07-12      &    58676.328225  &   1593  &  2.33 $\pm$ 0.15 & 7.11 $\pm$ 0.69 \\
2019-07-17      &    58681.609499  &   1983  &  2.17 $\pm$ 0.15 & 4.83 $\pm$ 0.48 \\
2019-07-22      &    58686.722614  &   1885  &  1.45 $\pm$ 0.22 & 2.98 $\pm$ 0.55 \\
2019-12-13      &    58830.935420  &   2023  &  1.44 $\pm$ 0.13 & 7.85 $\pm$ 0.87 \\
2020-01-02      &    58850.318109  &   2495  &  1.33 $\pm$ 0.17 & 5.47 $\pm$ 0.75 \\
2020-01-05      &    58853.943061  &   1666  &  1.38 $\pm$ 0.27 & 4.09 $\pm$ 0.87 \\
2020-01-08      &    58856.665569  &   1880  &  1.69 $\pm$ 0.20 & 3.73 $\pm$ 0.55 \\
2020-01-11      &    58859.830600  &   1983  &  1.81 $\pm$ 0.17 & 5.54 $\pm$ 0.70 \\
2020-01-19      &    58867.533771  &   1626  &  1.87 $\pm$ 0.15 & 5.88 $\pm$ 0.65 \\
2020-01-21      &    58870.011577  &   1546  &  1.94 $\pm$ 0.16 & 6.00 $\pm$ 0.65 \\
2020-01-24      &    58872.175468  &   1783  &  1.84 $\pm$ 0.14 & 6.57 $\pm$ 0.65 \\
2020-01-25      &    58873.171868  &   1798  &  2.10 $\pm$ 0.14 & 7.12 $\pm$ 0.69 \\
2020-01-27      &    58875.267506  &   1806  &  1.73 $\pm$ 0.19 & 4.42 $\pm$ 0.61 \\
2020-01-28      &    58876.193806  &   2035  &  1.47 $\pm$ 0.18 & 4.80 $\pm$ 0.67 \\
2020-01-31      &    58879.685875  &   1231  &  1.68 $\pm$ 0.19 & 5.88 $\pm$ 0.83 \\
2020-02-01      &    58880.114444  &   1681  &  1.58 $\pm$ 0.16 & 5.87 $\pm$ 0.77 \\
2020-02-05      &    58884.201846  &   2025  &  1.49 $\pm$ 0.15 & 6.30 $\pm$ 0.75 \\
2020-02-10      &    58889.473095  &   2417  &  1.57 $\pm$ 0.18 & 3.50 $\pm$ 0.48 \\
%2020-02-18      &    58897.609904  &   1808  &  1.44 $\pm$ 0.28 & 2.52 $\pm$ 0.56 \\
%2020-02-19      &    58898.600887  &   2138  &  1.58 $\pm$ 0.23 & 2.40 $\pm$ 0.43 \\
%2020-02-22      &    58901.339947  &   2340  &  1.21 $\pm$ 0.21 & 3.59 $\pm$ 0.67 \\
\hline
\end{tabular}                
\end{center}
\caption{Log and fitting results of {\em Swift}-XRT observations of B2 1420$+$32 using a PL model. 
%with $N_{\rm H}$ fixed to Galactic absorption,  i.e. 1.08$\times$10$^{20}$ cm$^{-2}$. 
Fluxes are corrected for the Galactic absorption.} 
\label{XRT_1424}
\end{table*}

\subsection{Long-term behaviour}\label{app:longterm}

\begin{figure*}[t!]
  \centering
  \includegraphics[width=0.9\textwidth]{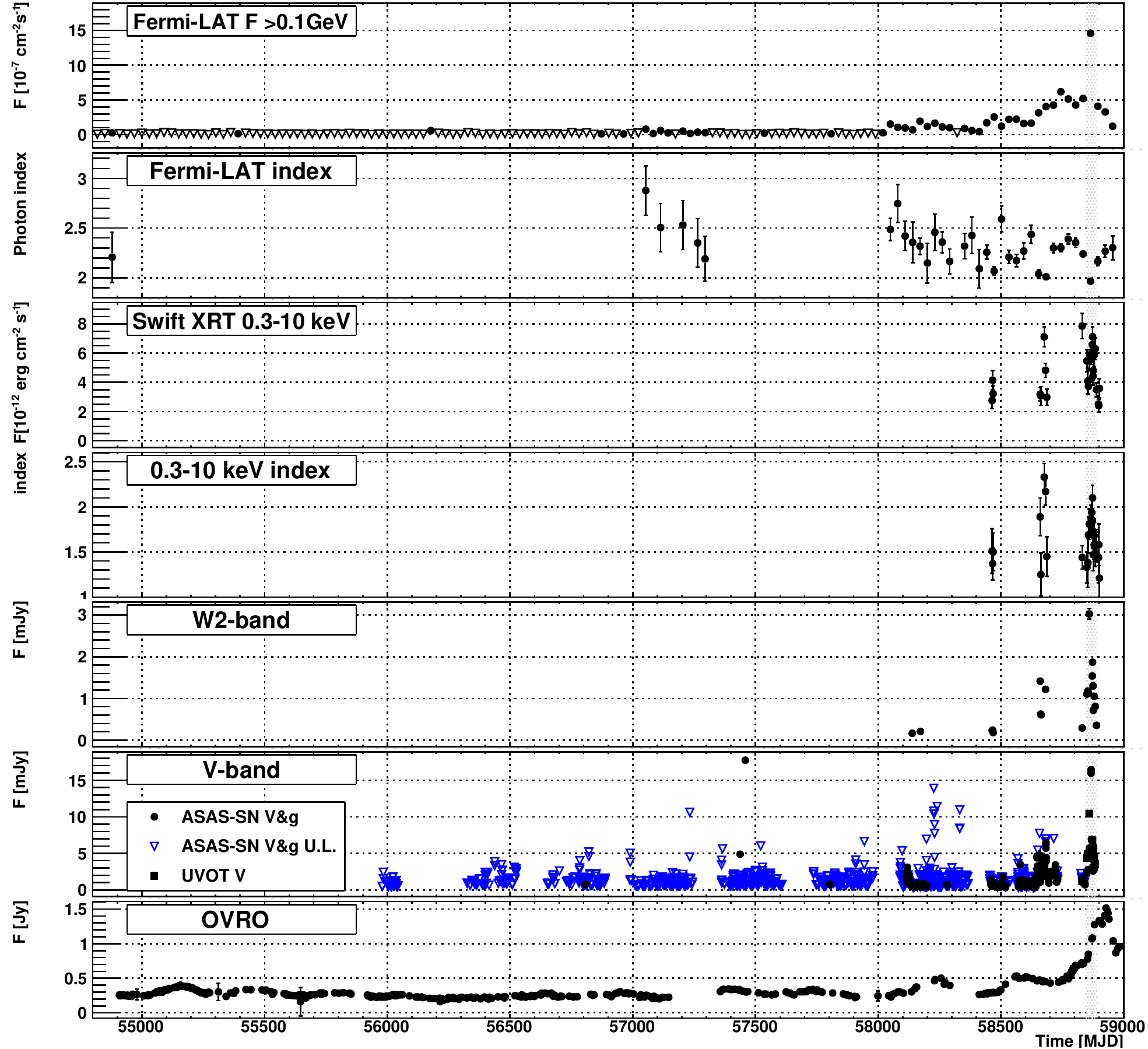}
  \caption{
     Long-term MWL light curve of \src\ (see titles and legends of individual panels). 
     %Optical observations are corrected for the Galactic attenuation. 
     The gray-shaded region shows the flaring period in the beginning of 2020. 
     \label{fig:mwl_lc_longterm}
}
\end{figure*}

In Fig.~\ref{fig:mwl_lc_longterm} we report the monitoring observations of \src\ in order to put the flaring state of January-February 2020 in the context of the long-term behaviour of the source. 
It is clear that the flaring period has been unique in the \fermi\ dataset of this source.
Similarly, the radio flux during the flare is also unique compared to previous measurements. 
On the other hand there is a prior report of a similar magnitude optical flare on 2016 March 11 (MJD=57458.5) in the ASAS-SN monitoring data \citep{st17}, however without a HE counterpart. 
Past X-ray data are unfortunately too sparse (and biased by Target of Opportunity observations) to judge about the typical behaviour of the source.

%%%%%%%%%%%%%%%%%%%%%%%%%%%%%%%%%%%%%%%%%%%%%%%%%%

\end{document}